\documentclass[12pt]{iopart}
\usepackage{iopams}
\usepackage{graphicx}
\usepackage{caption}
\usepackage{subcaption}
\usepackage{booktabs} 
\usepackage{multirow}
\usepackage{makecell}
\usepackage[table]{xcolor}

\usepackage{longtable}

\begin{document}

\title[TCV-X21 Hermes-3]{Validation of Hermes-3 turbulence simulations against the TCV-X21 diverted L-mode reference case}

\author{B D Dudson$^1$, M Kryjak$^{2,3}$, H Muhammed$^{2,3}$, J Omotani$^3$}

\address{$^1$ Lawrence Livermore National Laboratory, 7000 East Avenue, Livermore CA 94550, USA}
\address{$^2$ School of Physics, Engineering and Technology University of York, Heslington, York YO10 5DD, UK}
\address{$^3$ United Kingdom Atomic Energy Authority, Culham Center for Fusion Energy, Culham Science Centre, Abingdon, OX14 3DB, UK}

\begin{abstract}
  Electrostatic flux-driven turbulence simulations with the Hermes-3
  code are performed in TCV L-mode conditions in forward and reversed
  toroidal field configurations, and compared to the TCV-X21 reference
  dataset [D.S. Oliveira and T. Body {\it et al.} 2022] qualitatively and with a
  quantitative methodology. Using only the magnetic equilibrium, total
  power across the separatrix ($120$kW) and total particle flux to the
  targets ($3\times 10^{21}$/s) as inputs, the simulations produce
  time-averaged plasma profiles in good agreement with
  experiment. Shifts in the target peak location when the toroidal
  field direction is reversed are reproduced in simulation, including
  the experimentally observed splitting of the outer strike point into
  two density peaks.

  Differences between simulation and experiment include density
  profiles inside the separatrix and at the inner target in forward
  (favorable $\nabla B$) field configuration. These differences in
  target temperature in forward field configuration lead to
  differences in the balance of current to the inner and outer
  divertor in the private flux region.  The cause of these differences
  is most likely the lack of neutral gas in these simulations,
  indicating that even in low recycling regimes neutral gas plays an
  important role in determining edge plasma profiles.  These
  conclusions are consistent with findings in [D.S. Oliveira and T. Body {\it et
      al.} 2022].
\end{abstract}

\vspace{2pc}

\submitto{\NF}

\section{Introduction}

The edge and divertor region of tokamak plasmas consists of the
outermost part of the closed magnetic flux surfaces of the plasma
core, and open magnetic field lines that intersect material
surfaces. It is a transition region in which the flow of heat from the
plasma core meets neutral gas and solid surfaces, and as such has a
strong influence on both plasma performance and plasma facing
component lifetime. Understanding, predicting and controlling these
flows of heat is a long-standing challenge~\cite{PITTS2019100696} that
requires a combination of multi-channel diagnostics, novel
experiments, and sophisticated simulation models to
address~\cite{SCHWANDER2024106141, stangeby2000}.

Validation of plasma models is required to assess their reliability
when applied to interpretation of current experiments or predictions
for future devices. The TCV-X21 reference
dataset~\cite{Oliveira_2022}, consisting of experimental data from the
Tokamak \`a Configuration Variable (TCV~\cite{Duval_2024}), has been
published~\cite{tcv_x21_dataset} and compared against drift-fluid
simulations with GBS~\cite{ricci2012,Halpern2016},
GRILLIX~\cite{stegmeir_2019,Zholobenko_2021},
TOKAM3X~\cite{tamain2016,Nespoli_2019} and
SOLEDGE3X~\cite{BUFFERAND2024101824}, and the gyrokinetic code
GENE-X~\cite{Ulbl_2023}.  The results of these studies are guiding
rapid improvements in models of the boundary of magnetically confined
fusion plasmas.

In this study we use the TCV-X21 dataset to validate the Hermes-3
code~\cite{hermes-3-paper, dudson:hermes3} that is built on
BOUT++~\cite{dudson2015}. Earlier versions of this code were compared
to ISTTOK~\cite{Dudson_2021} Scrape-Off Layer (SOL) turbulence data
but significant changes to the physical model and implementation have
been made since then, including the addition of an ion pressure
equation.  The following section~\ref{sec:model} describes the
physical model and boundary conditions; results in forward and
reversed magnetic field configuration are presented in
section~\ref{sec:results}. We find differences in target conditions in forward
magnetic field configuration, and focus on analysis of the currents
flowing in the open field lines and through the sheath in
section~\ref{sec:currents}. Divertor heat fluxes, and flows of energy
in the simulation are studied in section~\ref{sec:heat_fluxes};
Quantitative validation metrics are summarised in
section~\ref{sec:validation}; Conclusions are summarised in
section~\ref{sec:conclusions}.

\section{Flux-driven turbulence model}
\label{sec:model}

Hermes-3~\cite{hermes-3-paper,dudson:hermes3} is capable of solving
for multiple ion and neutral species, in 1-, 2- or 3-D domains to
study steady-state, transient or turbulent phenomena. It is run here
as a 6-field flux-driven turbulence model with a single ion
species. Drift-reduced plasma fluid equations are solved for the
electron density ($n_e$), electron and ion pressures ($p_e$, $p_i$)
and electron and ion parallel momentum ($m_en_ev_{||e}$,
$m_in_iv_{||i}$), and vorticity $\omega$.  These equations are derived
from~\cite{simakov-2003, catto-2004, simakov-2004}, modified as
described in~\ref{apx:modifications}. Drift-reduced fluid
equations~\cite{Mikhailovskii_1971} are widely used in tokamak edge
modeling and have a wide variety of complexity and
implementations. Recently further improvements have been made in the
formulation of these models~\cite{halpern2023}, that are not
incorporated into the equations used here.

The electron fluid equations for density $n_e$, parallel momentum
$m_en_ev_{||e}$, and pressure $p_e = en_eT_e$ (where electron
temperature $T_e$ is in eV so that $eT_e$ has units of Joules, and all
other quantities are in SI units) are:
\begin{eqnarray}
  \frac{\partial n_e}{\partial t} &=& -\nabla\cdot\left[n_e \left(\mathbf{v}_{E} + \mathbf{b}v_{||e} + \mathbf{v}_{mag,e}\right)\right] + S_n \\
  \frac{\partial}{\partial t}\left(m_en_ev_{||e}\right) &=& -\nabla\cdot\left[m_en_ev_{||e} \left(\mathbf{v}_{E} + \mathbf{b}v_{||e} + \mathbf{v}_{mag,e}\right)\right] - \mathbf{b}\cdot\nabla p_e \nonumber \\
  &&- en_eE_{||} + F_{ei} \\
  \frac{\partial}{\partial t}\left(\frac{3}{2}p_e\right) &=& -\nabla\cdot\left[\frac{3}{2}p_e \left(\mathbf{v}_{E} + \mathbf{b}v_{||e}\right) + \frac{5}{2}p_e\mathbf{v}_{mag,e}\right] - p_e\nabla\cdot\left(\mathbf{v}_{E} + \mathbf{b}v_{||e}\right) \nonumber \\
  && + \nabla\cdot\left(\kappa_{||e}\mathbf{b}\mathbf{b}\cdot\nabla T_e\right) + S_{Ee} + W_{ei}
\end{eqnarray}
Where the electrostatic approximation is made, so that $E_{||} = -\mathbf{b}\cdot\nabla\phi$.
Cross-field drift velocities are the $E\times B$ drift $\mathbf{v}_{E} = \frac{\mathbf{b}\times\nabla\phi}{B}$
and electron magnetic drift $\mathbf{v}_{mag,e} = -T_e\nabla\times\frac{\mathbf{b}}{B}$.
The magnetic drift formulation
is analytically equivalent to the diamagnetic drift form used in~\cite{simakov-2004}, but is easier
to implement numerically in terms of conservative fluxes between cells~\cite{tamain2016}. $S_n$ is the external source
of particles; $F_{ei}$ is the collisional (Braginskii~\cite{braginskii1965}) friction between ions and electrons; $S_{Ee}$
is an external source of power to the electrons, and $W_{ei}$ is the collisional (Braginskii)
exchange of heat between ions and electrons. Details of the collisional terms implemented are given in~\cite{hermes-3-paper}.

The ion fluid equations assume quasineutrality so that $n_i = n_e$, and evolve
the ion parallel momentum $m_in_iv_{||i}$ and pressure $p_i$:
\begin{eqnarray}
  \frac{\partial}{\partial t}\left(m_in_iv_{||i}\right) &=& -\nabla\cdot\left[m_in_iv_{||i} \left(\mathbf{v}_{E} + \mathbf{b}v_{||i} + \mathbf{v}_{mag,i}\right)\right] - \mathbf{b}\cdot\nabla p_i \nonumber \\
  &&+ Z_ien_iE_{||} - F_{ei} \\
  \frac{\partial}{\partial t}\left(\frac{3}{2}p_i\right) &=& -\nabla\cdot\left[\frac{3}{2}p_i \left(\mathbf{v}_{E} + \mathbf{b}v_{||i}\right) + \frac{5}{2}p_i\mathbf{v}_{mag,i}\right] - p_i\nabla\cdot\left(\mathbf{v}_{E} + \mathbf{b}v_{||i}\right) \nonumber \\
  && + \nabla\cdot\left(\kappa_{||i}\mathbf{b}\mathbf{b}\cdot\nabla T_i\right) + S_{Ei} + S_n\frac{1}{2}m_in_iv_{||i}^2 - W_{ei} \nonumber \\
  && + \frac{p_i}{en_0}\nabla\cdot\left(\mathbf{J}_{||} + \mathbf{J}_d\right) \label{eq:ion_pressure}
\end{eqnarray}
Where the ion magnetic drift is $\mathbf{v}_{mag,i} = T_i\nabla\times\frac{\mathbf{b}}{B}$. The final term
in the ion pressure equation (\ref{eq:ion_pressure}) is not in the original equations~\cite{simakov-2004}
but is required for energy conservation. See \ref{apx:ion_energy} for details.

The vorticity $\omega$ can be derived from the divergence of the ion
polarisation current, and is implemented in Hermes-3 as:
\begin{equation}
  \omega = \nabla\cdot\left[\frac{m_in_0}{B^2}\nabla_\perp\left(\phi + \frac{p_i}{n_0}\right)\right] \label{eq:vorticity_definition}
\end{equation}
where $\nabla_\perp \equiv \nabla - \mathbf{b}\mathbf{b}\cdot\nabla$
and the Oberbeck-Boussinesq approximation~\cite{oberbeck1879} is made,
replacing the density in the polarisation current with a constant
$n_0$. The evolution of the vorticity is derived from current
continuity, such that the divergence of the sum of all currents is
zero: Polarisation current, parallel current $J_{||}$ and diamagnetic
current $J_d$:
\begin{eqnarray}
  \frac{\partial \omega}{\partial t} &=& -\nabla\cdot\left[\frac{m_i}{2B^2}\nabla_\perp\left(\mathbf{v}_E \cdot\nabla p_i\right) + \frac{\omega}{2}\mathbf{v}_E + \frac{m_in_0}{2B^2}\nabla_\perp^2\phi\left(\mathbf{v}_E + \frac{\mathbf{b}}{n_0B}\times\nabla p_i\right)\right] \nonumber \\
  &&+ \nabla\cdot\left(\mathbf{J}_{||} + \mathbf{J}_d\right) \label{eq:vorticity}
\end{eqnarray}
Where the divergence of the diamagnetic current is:
\begin{equation}
  \nabla\cdot\mathbf{J}_d = \nabla\cdot\left[\left(p_e + p_i\right)\nabla\times\frac{\mathbf{b}}{B}\right]
\end{equation}
The vorticity equation~(\ref{eq:vorticity}) has been modified to improve energy conservation in the model,
as described in \ref{apx:vorticity}.

The ion viscosity derived in~\cite{simakov-2003, catto-2004, simakov-2004}
is not included in these simulations, and does not include the
neoclassical correction recently implemented in
GRILLIX~\cite{zholobenko2024}, that replaces $\tau_i$ with a
combination of $\tau_i$ and ion bounce time. We plan to implement
similar improvements in future work that targets lower collisionality
regimes than are simulated here.

\subsection{Boundary conditions}

In the radial direction Neumann (zero gradient) boundary conditions
are applied to plasma density and pressures on both inner (core) and
outer (scrape-off layer) boundaries. A Dirichlet (zero value) boundary
is applied to the parallel flow of both electrons and ions at both radial
boundaries.

The calculation of the electrostatic potential $\phi$ from vorticity
$\omega$ is by inverting an elliptic
equation~(\ref{eq:vorticity_definition}) at every time step
iteration. The linear solve is performed in toroidal planes, invoking
the flute approximation to drop poloidal derivatives that couple
toroidal planes, as is usually done in BOUT++
models~\cite{dudson2015}.  We wish to impose zero-gradient boundary
conditions on $\phi$ at both boundaries, to allow poloidal variation
of the potential around the SOL and to avoid constraining the
potential at the core boundary. This would lead to an ill-posed linear
solve due to the null space of the operator.  Instead, we use a method
adapted from STORM~\cite{ukaea:storm, Riva_2019}: Dirichlet boundary
conditions are used in the linear solve for $\phi$, so that the
inversion is well posed, but the value of the boundaries relaxes
towards zero-gradient on a short timescale, here chosen to be $1\mu
s$.

At the sheath (target) boundaries Bohm-Chodura-Riemann boundary
conditions are applied~\cite{stangeby-1995}, using the multi-ion
magnetized sheath condition derived in~\cite{tskhakaya2005}. This
imposes sonic ion flow into the sheath, with a sheath heat transmission
that depends on the potential $\phi$. The potential at the sheath entrance is
not fixed; instead currents flow into or out of the sheath until a
(quasi-) steady-state potential is reached. This can lead to quite
complex patterns of plasma currents closing through the sheath that
are analysed in section~\ref{sec:currents}.

\subsection{Sources and geometry}
\label{subsec:sources}

Hermes-3 simulations are flux-driven with prescribed
spatially-varying sources of particles, electron and ion heating.
There are no neutrals in the simulations shown here, so the
experimentally inferred target particle flux of $3\times 10^{21}$
particles per second~\cite{Oliveira_2022} is injected into the
simulation domain in a thin region near the core boundary.  $120$kW of
heating power is injected, divided equally between electrons and ions,
and distributed evenly in poloidal angle near the core boundary using
the same source profile as the particle source.  Heat and particles
flow to the divertor targets, where they are removed through the
sheath boundary conditions at the target plates.  In these simulations
there are no neutrals and so no particle recycling: All particles
arriving at the sheath are removed from the simulation.

Simulations are performed in tokamak geometry using the curvilinear
shifted metric coordinate system usually used in
BOUT++~\cite{dudson2015}. The tangential basis vectors are a radial
coordinate ($x$) in the direction of $\nabla\psi$, a poloidal
coordinate ($y$) that is aligned to the magnetic field direction, and
a toroidal angle coordinate ($z$). Aligning the poloidal coordinate
with the magnetic field enables a coarse poloidal resolution to be
used while resolving small-scale field-aligned turbulent structures,
but at the cost of introducing a coordinate singularity at the
X-point, so that the region around the X-point is poorly resolved in
the poloidal plane. Meshes are constructed using the Hypnotoad
tool~\cite{hypnotoad}.

Two TCV equilibria are studied here: A ``forward'' ($+$) toroidal
magnetic field configuration favorable $\nabla B$, with ion magnetic
drift $\mathbf{v}_{mag,i}$ downwards towards the X-point; and a
``reverse'' ($-$) toroidal field configuration in which the ion magnetic
drift is upwards.

\subsection{Numerical methods}

Conservative finite difference methods are used in Hermes-3, with
cross-field drifts as described in \cite{Dudson2017} and parallel
dynamics in \cite{hermes-3-paper,dudson2019}: All quantities are cell
centered, and 2$^{nd}$-order operators are formulated in terms of
fluxes between cells; The Monotonised Central (MC) limiter is used to
reconstruct cell edge values from cell center values. In the direction
along the magnetic field a Lax flux is used to stabilise grid-scale
oscillations, taking the electron thermal speed as the fastest wave in
the system. Time integration is fully implicit with an adaptive order
and timestep, implemented in the SUNDIALS CVODE
library~\cite{hindmarsh2005}.

Simulations shown here were performed using parameters shown in
table~\ref{tab:mesh}.
\begin{table}[h]
  \caption{Simulation mesh parameters}
  \label{tab:mesh}
  \centering
  \begin{tabular}{c c}
    \toprule
    Mesh size ($\psi$, $\theta$, $\zeta$) & $64 \times 32 \times 81$ \\
    Fraction of a torus & 1/5  \\
    Radial resolution [mm] & 0.3 - 7.6 \\
    Binormal resolution [mm] & 0.2 - 3.0 \\
    Sound gyro-radius [mm] & 0.1 - 1.4 \\
    \bottomrule
  \end{tabular}
\end{table}

\begin{figure}
  \centering
  \begin{subfigure}[b]{0.49\textwidth}
    \centering
    \includegraphics[width=\textwidth]{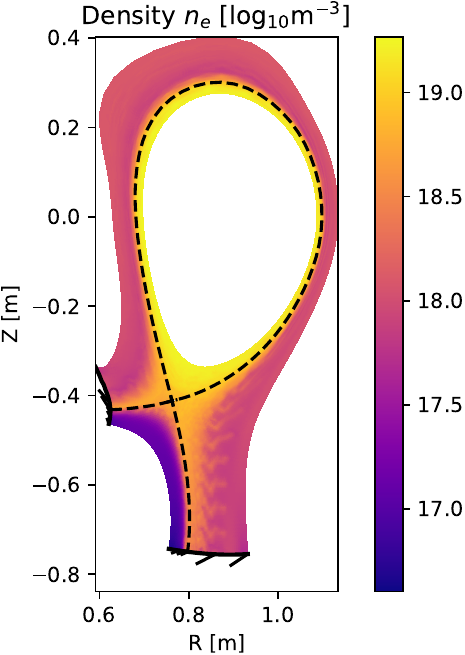}
    \caption{Snapshot of the electron density $n_e$ on a logarithmic scale.}
    \label{fig:density_log}
  \end{subfigure}
  \hfill
  \begin{subfigure}[b]{0.49\textwidth}
    \centering
    \includegraphics[width=\textwidth]{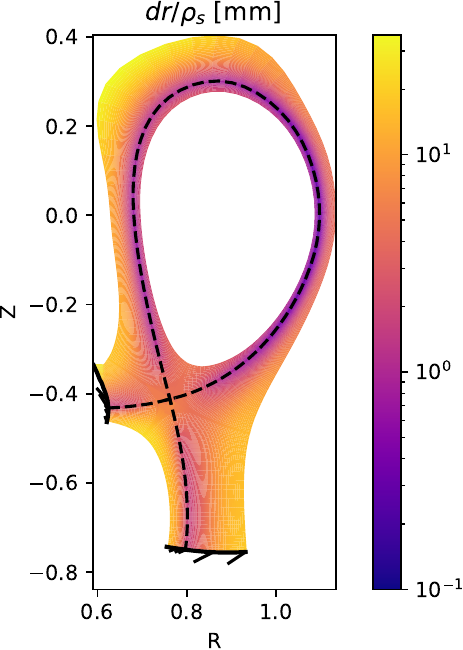}
    \caption{Radial grid cell spacing divided by ion sound gyroradius.
      Calculated using the simulation time-averaged temperature profiles.
      The yellow regions close to the separatrix are well resolved ($dr/\rho_s\simeq 1$).}
    \label{fig:dr_rho_s}
  \end{subfigure}
  \caption{Poloidal slices through the 3D domain in the forward field configuration.
    The simulations solve for an annulus around the separatrix, and do not solve for
    the core region.}
  \label{fig:forward_poloidal}
\end{figure}

The resolution of the mesh should be sufficient to resolve the plasma
``blob'' structures, typically on the scale of a few to tens of ion
sound gyroradii $\rho_s = \sqrt{eT_e/m_i}\left(m_i / eB\right)$. This
scale-length depends on the local plasma temperature, so must be
calculated in post-analysis.  Figure~\ref{fig:dr_rho_s} shows the
ratio of radial grid spacing $dr$ to $\rho_s$, using time-averaged temperature
profiles from the Hermes-3 simulation.
It can be seen that the turbulence scales are well resolved (light
yellow) close to the separatrix, in both closed field-line region and
divertor legs. In the outer scrape-off layer (SOL) and private flux region
(PFR) the electron temperature is low and the mesh is unable to
resolve the smallest scales. We therefore expect best agreement
between simulation and experiment in the near SOL in the following
sections.

\section{Results}
\label{sec:results}

In the following results the forward field simulation was run for
$7.3$ms, with results taken from the last $1.9$ms. The reversed
field simulation was run for $6.9$ms, with results taken from the last
$1.9$ms. The computational cost at this low resolution was
approximately $6\times 10^3$ core-hours per ms of simulation time,
running on 64 cores.

\subsection{Midplane profiles}

A summary of separatrix quantities from experimental
measurements~\cite{Oliveira_2022} and Hermes-3 simulations is given in
table~\ref{tab:separatrix}.
\begin{table}[h]
  \caption{Separatrix parameters at outboard midplane in forward toroidal field configuration}
  \label{tab:separatrix}
  \centering
  \begin{tabular}{c c c c}
    \toprule
    & Hermes-3 & TCV \\
    \midrule
    Separatrix $T_e$ [eV] & $39 \pm 3$ & $35 \pm 5$ \\
    $T_e$ width, $\lambda_T$ [cm] & $0.8 \pm 0.1$ & $1.0 \pm 0.4$ \\
    Separatrix $n_e$ [$10^{18}$m$^{-3}$] & $11.7 \pm 1.5$ & $6.7 \pm 0.8$ \\
    $n_e$ width, $\lambda_n$ [cm] & $0.5 \pm 0.1$ & $0.9 \pm 0.2$ \\
    \bottomrule
  \end{tabular}
\end{table}
The temperature and temperature scale-length in the near Scrape-Off
Layer (SOL) are quite well matched (within one standard deviation).
The Hermes-3 simulations however have separatrix densities that are
higher than experiment, by a factor of $\sim 1.8$. We hypothesize that
this difference in density is due to the different particle source
locations, since the total particle flux to the targets is
approximately matched: In experiment neutrals are ionized in the
divertor leg, whereas in these simulations the only source of
particles is in the closed field-line region.

\begin{figure}
  \centering
  \begin{subfigure}[b]{0.49\textwidth}
    \centering
    \includegraphics[width=\textwidth]{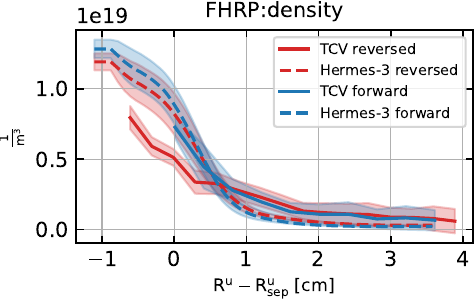}
    \caption{Electron density $n_e$. Simulations produce higher
      density than experiment. The shift to lower density in reversed
      configuration (red) as compared to forward configuration (blue)
      is reproduced.}
    \label{fig:midplane_density_rev}
  \end{subfigure}
  \hfill
  \begin{subfigure}[b]{0.49\textwidth}
    \centering
    \includegraphics[width=\textwidth]{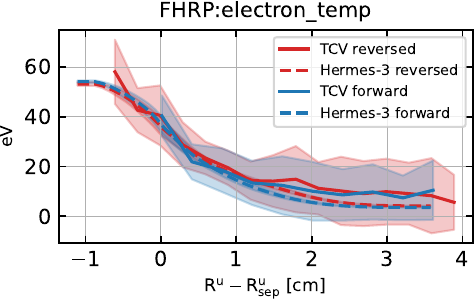}
    \caption{Electron temperature $T_e$. Mean profiles are well matched
      between experiment and simulation, but standard deviation is larger
      in experiment than simulation.}
    \label{fig:midplane_te_rev}
  \end{subfigure}
  \hfill
  \begin{subfigure}[b]{0.49\textwidth}
    \centering
    \includegraphics[width=\textwidth]{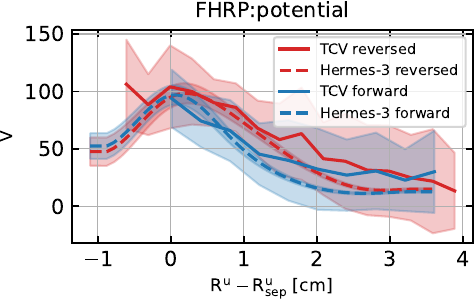}
    \caption{Potential $\phi$. The magnitude and radial shift with toroidal
      field direction is reproduced, but fluctuation standard deviations
      are smaller in simulation than experiment.}
    \label{fig:midplane_potential_rev}
  \end{subfigure}
  \caption{Midplane profiles. Forward field cases are shown as blue
    lines, and reversed field as red lines.  Experimental data are
    solid lines, and Hermes-3 simulations are dashed.}
  \label{fig:midplane_rev}
\end{figure}

Radial profiles of density $n_e$, electron temperature $T_e$ and
plasma potential $\phi$ are shown in
figure~\ref{fig:midplane_rev}. Solid lines show experimental data, and
dashed lines the Hermes-3 simulation. As previously, shaded regions
indicate the standard deviation of the data. Forward (favorable
$\nabla B$) configuration is plotted in blue, and reverse
configuration in red.

In both forward and reverse configuration the density profiles
(figure~\ref{fig:midplane_density_rev}) are higher in simulation than
experiment, as shown in table~\ref{tab:separatrix}. In both experiment
and simulation we see that the density is higher in forward-field
configuration (blue lines) than reversed (red).

The electron temperature is quite insensitive to the magnetic field
direction, both at the midplane (figure~\ref{fig:midplane_te_rev}) and
at the targets (figures~\ref{fig:lfs_te_rev} and
\ref{fig:hfs_te_rev}).  The simulation and experimental profiles are
within error bars (shaded region), but the simulation is
systematically cooler than experiment in the far SOL and has lower
fluctuation levels, shown as narrower shaded regions.

In both experiment and simulation the plasma potential
(figure~\ref{fig:midplane_potential_rev}) shifts with the change in
toroidal magnetic field, so that the potential in the SOL is higher
with reversed toroidal field than forward field. The temperature
profiles are largely unchanged, so this shift is likely to be
dominated by the diamagnetic drift direction rather than changes in
the sheath potential with electron temperature. At the low-field side
target (figure~\ref{fig:lfs_potential_rev}) the same shift in
potential is seen, accompanied by changes in time-averaged current
into the sheath (figure~\ref{fig:lfs_current_rev}), that will be
discussed in section~\ref{sec:currents}.

\subsection{Low-field side divertor target}

\begin{figure}
  \centering
  \begin{subfigure}[b]{0.49\textwidth}
    \centering
    \includegraphics[width=\textwidth]{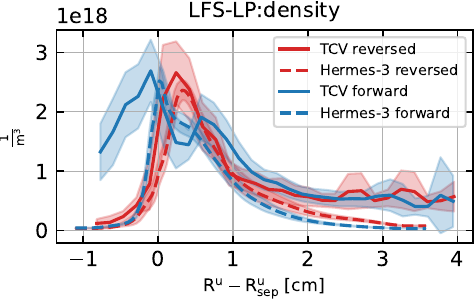}
    \caption{Electron density $n_e$}
    \label{fig:lfs_density_rev}
  \end{subfigure}
  \hfill
  \begin{subfigure}[b]{0.49\textwidth}
    \centering
    \includegraphics[width=\textwidth]{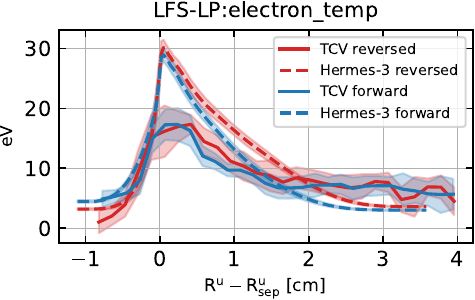}
    \caption{Electron temperature $T_e$}
    \label{fig:lfs_te_rev}
  \end{subfigure}
  \hfill
  \begin{subfigure}[b]{0.49\textwidth}
    \centering
    \includegraphics[width=\textwidth]{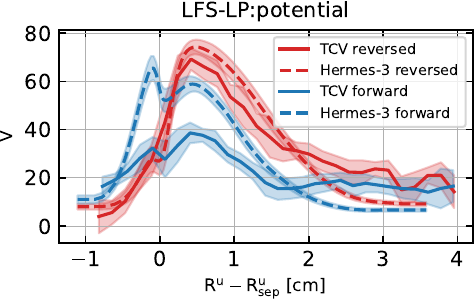}
    \caption{Plasma potential $\phi$}
    \label{fig:lfs_potential_rev}
  \end{subfigure}
  \begin{subfigure}[b]{0.49\textwidth}
    \centering
    \includegraphics[width=\textwidth]{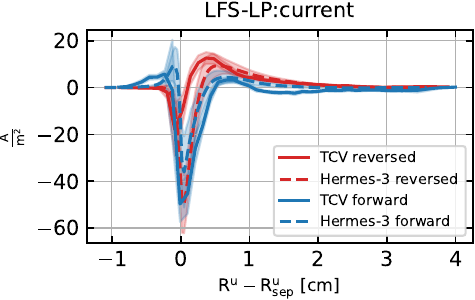}
    \caption{Parallel current $j_{||}$}
    \label{fig:lfs_current_rev}
  \end{subfigure}
  \caption{Low-Field Side (LFS) divertor profiles. Forward field cases
    are shown as blue lines, and reversed field as red lines.
    Experimental data from Langmuir probes are solid lines, and
    Hermes-3 simulations are dashed. All profiles are mapped along
    flux surfaces in $\psi$ space from divertor target to major radius
    $R^u$ at midplane.}
  \label{fig:lfs_rev}
\end{figure}

Density, temperature, potential and current profiles measured using
Langmuir probes at the low-field (outboard) divertor target are shown
in figure~\ref{fig:lfs_rev}. The radial coordinate $R_u - R_{sep}$ is
the distance from the separatrix at the outboard midplane, mapped in
poloidal flux $\psi$ from the divertor target. Locations in the PFR
are not connected to the outboard midplane, so the mapping is to the
flux surface in the core with the same $\psi$ value as the PFR flux
surface.
There is a qualitative change in the density profile
(fig~\ref{fig:lfs_density_rev}) between forward and reversed field
cases: In the reversed field case a single peak is seen in both
experiment and simulations, at a midplane-mapped location of $R^u -
R^u_{sep}\simeq 0.3$cm. When the toroidal field is in the forward
direction, Hermes-3 reproduces the experimentally observed broadening
of the density peak, sometimes splitting into two peaks like the
potential $\phi$, with the location of maximum density shifting
towards the separatrix. There is however a notable difference in the
density between simulation and experiment in the PFR of the
forward-field case.  This could be caused by stronger transport into
the PFR from the SOL in experiment, or due to ionization sources that
are present in experiment but not simulation.

As with the midplane profiles, the temperature profiles
(figure~\ref{fig:lfs_te_rev}) are quite insensitive to field
direction, particularly near the separatrix. Simulation temperatures
are more peaked, and systematically higher, than experiment.  Going
from forward (blue) to reversed (red) there is reduction in the PFR
temperature and increase in the SOL temperature, in both experiment
(solid) and simulation (dashed).

The plasma potential (figure~\ref{fig:lfs_potential_rev}) is well
matched in reverse field configuration (red lines), though Hermes-3
does not reproduce the relatively flat profiles of electron
temperature and potential in the far SOL.  This may be related to the
far SOL being under-resolved (figure~\ref{fig:dr_rho_s}). In the
forward field configuration (blue lines) both experiment and simulation
show two peaks in potential, though the PFR has a much higher
voltage in simulation than experiment. In this forward configuration
the PFR sheath is in the ion saturation regime, so that large changes in
potential $\phi$ cause only small changes in sheath current $j_{||}$
(figure~\ref{fig:lfs_current_rev}). In section~\ref{sec:currents}
the currents into the sheath are analysed in more detail.

\subsection{High-field side divertor target}
\label{sec:rev_hfs}

\begin{figure}
  \centering
  \begin{subfigure}[b]{0.49\textwidth}
    \centering
    \includegraphics[width=\textwidth]{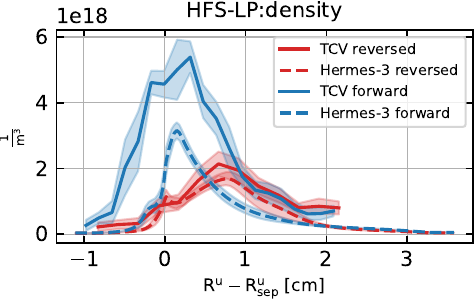}
    \caption{Electron density $n_e$. Simulations reproduce the observed
      shift in density peak, but in forward field the density peak is
      significantly broader in experiment than simulation.}
    \label{fig:hfs_density_rev}
  \end{subfigure}
  \hfill
  \begin{subfigure}[b]{0.49\textwidth}
    \centering
    \includegraphics[width=\textwidth]{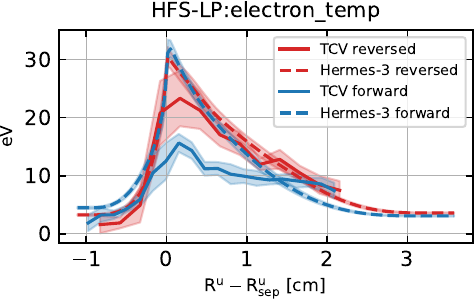}
    \caption{Electron temperature $T_e$. In forward field the HFS
      target is cooler in experiment (blue solid line) relative to
      simulation (blue dashed line) and reversed field (red lines).}
    \label{fig:hfs_te_rev}
  \end{subfigure}
  \hfill
  \begin{subfigure}[b]{0.49\textwidth}
    \centering
    \includegraphics[width=\textwidth]{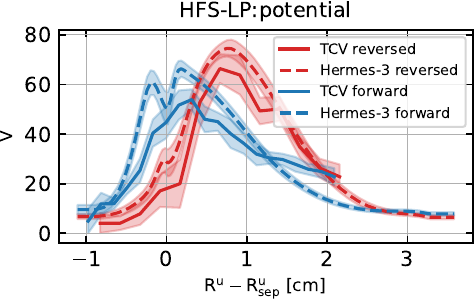}
    \caption{Plasma potential $\phi$. In forward field the potential is higher
    in simulation than experiment in the Private Flux region (PFR), $R^u - R^u_{sep} < 0$.}
    \label{fig:hfs_potential_rev}
  \end{subfigure}
  \begin{subfigure}[b]{0.49\textwidth}
    \centering
    \includegraphics[width=\textwidth]{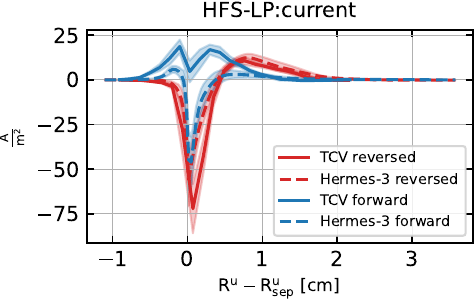}
    \caption{Parallel current $j_{||}$. In forward field the current in experiment
      has the opposite sign to experiment.}
    \label{fig:hfs_current_rev}
  \end{subfigure}
  \caption{High-Field Side (HFS) divertor profiles. Forward field
    cases are shown as blue lines, and reversed field as red lines.
    Experimental data from Langmuir probes are shown as solid lines,
    and Hermes-3 simulations are dashed lines. All profiles are mapped
    along flux surfaces in $\psi$ space from divertor target to major
    radius $R^u$ at midplane. Hermes-3 agrees well with experimental
    profiles in reversed field, but not in forward field.}
  \label{fig:hfs_rev}
\end{figure}

Plasma profiles at the inboard (high-field side, HFS) target
are shown in figure~\ref{fig:hfs_rev}.
Hermes-3 simulations are able to match well the density, electron
temperature, potential and parallel current profiles at the high-field
side target in the reversed field configuration (red lines in
figure~\ref{fig:hfs_rev}). Simulations reproduce features of the
target profiles including density peaking in the SOL ($R^u -
R^u_{sep}\simeq 0.8$cm), the location of the sign reversal in the
parallel current ($R^u - R^u_{sep}\simeq 0.5$cm), and both shape and
magnitude of the plasma potential profile.

Going from reversed field (red) to forward field (blue) case, the
density peak (figure~\ref{fig:hfs_density_rev}) shifts closer to the
separatrix in both simulation and experiment, but in experiment the
peak is significantly broader than in simulation. The temperature
(figure~\ref{fig:hfs_te_rev}) also shifts, but is significantly
lower in experient in the forward magnetic field than reverse field.

A striking difference between simulation and experiment is the
parallel current into the sheath (figure~\ref{fig:hfs_current_rev})
that is quite well matched between simulation and experiment in
reversed field, but different in forward field. The reversed field
case shows that if the density, temperature and potential
(figure~\ref{fig:hfs_potential_rev}) are comparable between experiment
and simulation, then the parallel current also agrees. This indicates
that the sheath boundary condition implemented in
Hermes-3~\cite{tskhakaya2005} is consistent with the sheath
characteristics assumed in the Langmuir probe analysis in this plasma
regime. In the next section we examine in more detail the currents in
forward and reversed magnetic field configurations, in order to
understand the observed differences.

\section{Sheath currents}
\label{sec:currents}

The currents in the SOL and PFR are determined by combination of
diamagnetic, polarisation and parallel currents, complicated by the
nonlinear and asymmetric response of the sheath to electrostatic
potential perturbations: In these simulations the wall is conducting
so that the current through the sheath into the
wall~\cite{tskhakaya2005} has the form of
equation~\ref{eq:sheath_current}.
\begin{equation}
  j_{sh} \simeq nc_s\left[1 - \exp\left(-\phi / T_e\right)\right]
  \label{eq:sheath_current}
\end{equation}
and goes to zero when the potential $\phi\simeq 2.8 T_e$ for a
deuterium plasma.  In general the current is non-zero, depending
nonlinearly on the local temperature, density and potential. The
balance of currents changes when the magnetic field is reversed,
because the magnetic drift (equivalently, diamagnetic current)
reverses direction.

In the forward field case we find good agreement in the parallel
current at the low field side (blue lines in
figure~\ref{fig:lfs_current_rev}), but not at the high-field side
(blue lines in figure~\ref{fig:hfs_current_rev}) where the Hermes-3
current is more negative (out of the target) than experiment. In
reversed field configurations we find good agreement at the high-field
side (red lines in figure~\ref{fig:hfs_current_rev}), with good
agreement in the density, electron temperature, and plasma potential,
particularly in the region $R^u - R^u_{sep}\simeq 0 - 0.5$cm. At the
low-field side, however, (red lines in
figure~\ref{fig:lfs_current_rev}) the simulation time-averaged current
is more negative than measured experimentally.

\subsection{Currents in the Private Flux Region}
\label{eq:pfr_currents}

\begin{figure}
  \centering
  \begin{subfigure}[t]{0.49\textwidth}
    \centering
    \includegraphics[width=0.7\textwidth]{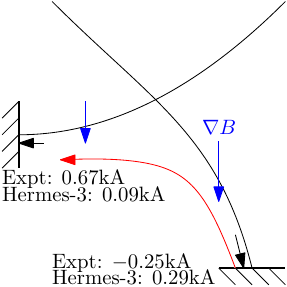}
    \caption{Forward field. Downward $\nabla B$ currents (blue arrows)
      are balanced by Plate Closing Currents (PCC) into the sheath
      (black arrows). In experiment an additional thermoelectric
      current (red arrow) flows from outer to inner target, reversing
      the current at the outer (LFS) target.}
    \label{fig:pf_currents_fwd}
  \end{subfigure}
  \hfill
  \begin{subfigure}[t]{0.49\textwidth}
    \centering
    \includegraphics[width=0.7\textwidth]{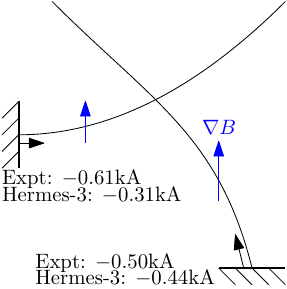}
    \caption{Reverse field. Upward $\nabla B$ currents (blue arrows)
      are balanced by Plate Closing Currents
      (PCC~\cite{rozhansky2020}) out of the sheath (black arrows).}
    \label{fig:pf_currents_rev}
  \end{subfigure}
  \caption{Currents into and out of the Private Flux Region (PFR), in
    (a) forward (favorable $\nabla B$) and (b) reverse (unfavorable $\nabla
    B$) toroidal field configurations.}
  \label{fig:pf_currents}
\end{figure}

The analysis of currents into the sheaths is simplified by considering
the PFR, because the magnetic topology isolates it from the
core plasma dynamics. The currents into and out of the PFR are sketched in
figure~\ref{fig:pf_currents} for both forward and reversed field.
In reverse (unfavorable) configuration, ion $\nabla B$ and curvature
drifts drive a current from the PFR upwards into the SOL. This
is balanced by Plate Closing Currents (PCC) from the plasma into the
sheath~\cite{rozhansky2020}, illustrated in figure~\ref{fig:pf_currents_rev}.
The total current into both targets is
similar between experiment ($-1.1$kA) and Hermes-3 simulation
($-0.75$kA), and is divided approximately equally between inner (HFS)
and outer (LFS) targets: At HFS target the experimentally measured
current at in the PFR is $-0.61$kA, simulated is $-0.31$kA; at
LFS the experimentally measured current is $-0.50$kA, and simulated is
$-0.44$kA.

In forward (favorable) configuration, ion $\nabla B$ and curvature
drifts drive a current from the SOL into the PFR, so that net current
into the PFR reverses sign relative to the reverse configuration. As
in reversed field configuration, the magnitude and direction of the
total current to the targets is similar: experiment $0.42$kA,
simulation $0.38$kA. In this case the currents at the two targets are
quite different between experiment and simulation. At the high-field
side (HFS) target the experiment measured $0.67$kA into the PFR, but
simulation only $0.09$kA; At the low-field side (LFS) the
experimentally measured $-0.25$kA is opposite in sign to the $0.29$kA
predicted in simulation.

We attribute this difference to a strong thermoelectric current from
outer (LFS) to inner (HFS) targets in the PFR~\cite{staebler1989}
that is present the experiment but not
simulation, due to differences in target temperatures.  As discussed
in section~\ref{sec:rev_hfs}, the temperature at the inner (HFS)
target in forward field configuration (figure~\ref{fig:hfs_te_rev}) is
significantly cooler in experiment than simulation, perhaps due to
cooling by neutral recycling that is present in experiment but not
simulation.  The current into the sheath is approximately given by
equation~\ref{eq:sheath_current}. At the LFS the potential is less
than $2.8T_e$ ($T_e\sim 15$eV, $\phi\sim 30$V), whereas at HFS the
potential is above this threshold ($T_e \sim 12$eV, $\phi\sim
40$V). The lower temperature at HFS therefore drives a positive
current into the HFS target, drawn from the higher temperature LFS
target.

In summary, in both forward and reverse field cases the total current
to both targets is similar between experiment and simulation, being
determined by the magnetic drift current across the separatrix. The
balance of currents between the target plates shows differences
that correlate to cooling of the inner target. This is likely to be
due to plasma-neutral interactions that are not included in these
simulations.

\section{Divertor heat fluxes and power balance}
\label{sec:heat_fluxes}

The heat flux through the sheath and into the divertor targets due to
electrons and ions is the heat flux $q_{||}$ along the magnetic field:
\begin{equation}
  q_{||e,i} = \gamma_{e,i}en_{e,i}T_{e,i}C_i \qquad .
\end{equation}
The electron and ion sheath heat transmission factors,
$\gamma_e$ and $\gamma_i$, are calculated for a single ion species
as~\cite{tskhakaya2005}:
\begin{eqnarray}
  \gamma_e &=& 2 + \left(\phi_{sheath} - \phi_{wall}\right) / T_{e,sheath} \\
  \gamma_i &=& 2.5 + \frac{1}{2}M_iC_i^2 / eT_{i,sheath}
\end{eqnarray}
Where the ion sheath velocity $C_i^2 = e\left(5/3 T_i +
T_e\right)/M_i$.  In these simulations $\gamma_e$ varies between $3.5$
and $9$ over time and space, with time-averaged values varying in
space between $4$ and $7.5$. The time-averaged ion heat transmission
factor, $\gamma_i$ varies between $3.5$ and $4.5$.

\begin{figure}
  \centering
  \begin{subfigure}[t]{0.49\textwidth}
    \centering
    \includegraphics[width=\textwidth]{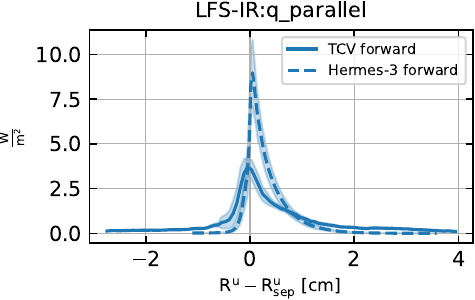}
    \caption{Forward field}
    \label{fig:q_parallel_fwd}
  \end{subfigure}
  \hfill
  \begin{subfigure}[t]{0.49\textwidth}
    \centering
    \includegraphics[width=\textwidth]{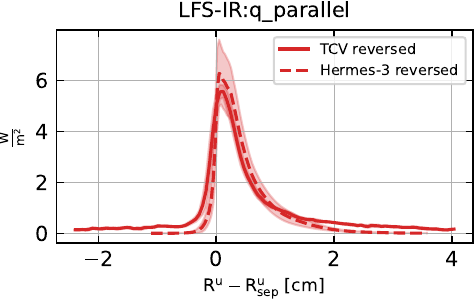}
    \caption{Reverse field}
    \label{fig:q_parallel_rev}
  \end{subfigure}
  \caption{Solid lines: Parallel heat flux at the outboard (LFS) target as measured by infrared (IR) camera diagnostic, mapped in poloidal flux space to the outboard midplane. Dashed lines: Hermes-3 simulation. Shaded areas in the simulation data indicate a standard deviation of the heat flux, over time and toroidal angle.
  }
  \label{fig:q_parallel}
\end{figure}

Experimentally the heat flux is measured using an infrared (IR)
camera~\cite{Oliveira_2022}, with target surface heat flux mapped to
parallel heat flux. Results are shown in figure~\ref{fig:q_parallel}
for forward and reverse field configurations.
Note that the simulation heat fluxes do not include surface
recombination energy that is present in experiment.  Hermes-3
simulations produce a higher peak power density at the LFS divertor
but a lower total power to the outboard target than
experiment. Possible explanations for the additional broadening in
experiment include differences in plasma turbulence, radiated power,
and other neutral interactions.

Integrating heat fluxes across the outboard (LFS) divertor in
figure~\ref{fig:q_parallel} provides total powers shown in
table~\ref{tab:heat_fluxes}. Only LFS target heat flux is measured by
the IR camera, but the simulation heat fluxes at both targets are
given.
\begin{table}[h]
  \caption{Power in kW to outboard (LFS) and inboard (HFS) divertor
    targets in forward (+) and reverse (-) configurations. TCV
    experimental results at LFS are obtained by integrating infrared
    (IR) camera measurements (figure~\ref{fig:q_parallel}). Simulation
    results do not include surface recombination power.}
  \label{tab:heat_fluxes}
  \centering
  \begin{tabular}{c l c c c c}
    \toprule
    & & Hermes-3 (+) & TCV (+) & Hermes-3 (-) & TCV (-) \\
    \midrule
    LFS & Total & 39 & 36.5 & 40 & 46.5 \\
    & Electron & 21 & & 21 & \\
    & Ion & 18 & & 19 & \\
    \midrule
    HFS & Total & 37 & & 30 & \\
        & Electron & 20 & & 14 & \\
        & Ion & 17 & & 16 & \\
    \bottomrule
  \end{tabular}
\end{table}
Hermes-3 simulations find that in forward (+) configuration the power
balance is approximately equal between LFS and HFS divertor targets,
while in reverse (-) configuration more power goes to the LFS target
than HFS.

\begin{figure}
  \centering
  \includegraphics[width=\textwidth]{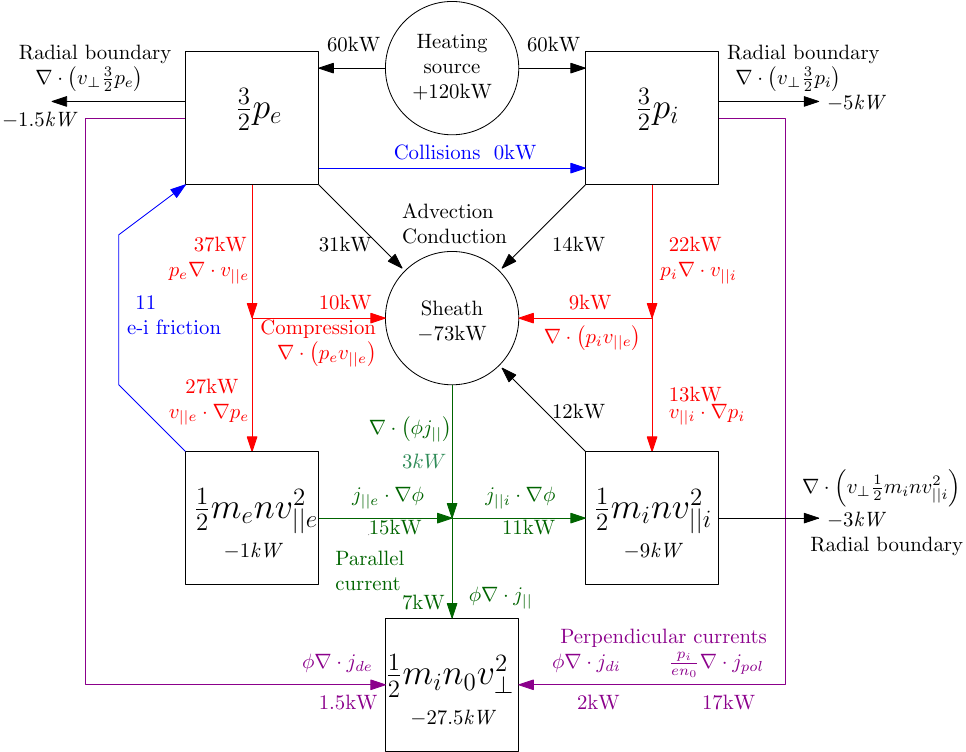}
  \caption{Forward field energy flow analysis. Energy flows from the
    heating source into thermal energy ($\frac{3}{2}p_{e,i}$) and from
    there to perpendicular kinetic energy
    ($\frac{1}{2}m_in_0v_\perp^2$), parallel electron and ion kinetic
    energy ($\frac{1}{2}m_{e,i}n_0v_{||e,i}^2$), before being
    deposited in the sheath, transported to radial boundaries, or lost
    in numerical dissipation. Positive numbers indicate heat sources
    and transfer channels; negative numbers indicate energy sinks. The
    heating source and sheath sink are shown as circles; Energy forms
    as square boxes, with negative numbers indicating the energy
    imbalance (if any). Energy transfer channels are labeled with
    approximate energy flow in kW, and are colored to indicate
    parallel advection and conduction (black); parallel compression
    (red); parallel currents (green); perpendicular currents
    (magenta); and collisions (blue). See~\ref{apx:energy} for
    details.}
  \label{fig:energy_flow}
\end{figure}

As described in section~\ref{subsec:sources}, both forward and reverse
simulations are driven by $120$kW of heating power inside the
separatrix, $60$kW into each species. We should therefore expect a
similar power to arrive at the divertor targets.  To understand the
discrepancy between input power and power arriving at the target
plates in these simulations, we have instrumented Hermes-3 with
diagnostics to track energy flows and transfer channels that are
described in \ref{apx:energy}. Taking the forward field case,
integrating energy transfer channels over the domain and averaging
over time, we obtain results shown in figure~\ref{fig:energy_flow}.

The energy flows in figure~\ref{fig:energy_flow} are averaged over the
domain and a period of $78\mu$s. The system has reached quasi-steady
state, such that the rate of change of the thermal energy in the
system is less than $0.1$kW over this averaging period.  Nevertheless,
the energy fluxes given here have uncertainties and should be
considered approximate: The Root-Mean-Square (RMS) value of the energy
transfer channels are orders of magnitude larger than their
averages. For example, the transfer $\frac{p_i}{en_0}\nabla\cdot
j_{pol}$ between ion thermal and perpendicular kinetic energy has an
RMS value of around $0.7$MW/m$^3$ and peak values (positive and
negative) of order $20$MW/m$^3$. The volume of the domain is
$1.3$m$^3$, so fluctuations are significant as compared to the mean
power flows through the system. We therefore do not expect exact
balance between energy flows over time averaging periods shorter than
the energy confinement time.  The analysis here has been
repeated for averages over short time windows of $0.1$ms and $0.2$ms,
finding qualitatively the same results.

Electron and ion thermal energy are conserved well: Each have $60$kW
of heating power that balances losses to within $1$kW. This
conservation is expected because species pressures are evolved by
exchanging fluxes between cells. Of the $120$kW input power,
approximately $6.5$kW is transported as thermal energy to the radial
domain boundaries (walls).

Electron and ion parallel kinetic energy
($\frac{1}{2}m_{e}n_0v_{||e}^2$ and $\frac{1}{2}m_{i}n_0v_{||i}^2$)
are not evolving variables, but are derived from evolution of momentum
and density. Electron parallel kinetic energy is driven by the
parallel electron pressure gradient ($27$kW), dissipated by
electron-ion friction ($11$kW) and parallel electric fields
($15$kW). Approximately $1$kW is unaccounted for in the electron
parallel kinetic energy balance, and is likely being lost as numerical
dissipation. Similar analysis of the ion parallel kinetic energy finds
approximately $9$kW imbalance.  In the transfer terms involving
parallel currents $j_{||}$ and electric potential $\phi$ (colored
green in figure~\ref{fig:energy_flow}) there is an energy flow $\int
\phi j_{||}\cdot dS$ of order $3$kW out of the sheath.  This energy
flow accounts for the difference between the total power to the
divertor targets given in table~\ref{tab:heat_fluxes} for forward
field ($76$kW), and the total power to the sheath in
figure~\ref{fig:energy_flow} ($73$kW).

The ion perpendicular flow ($\frac{1}{2}m_{i}n_0v_{\perp}^2$) is also
a derived quantity, being related to the vorticity equation
(see~\ref{apx:energy}). We observe a transfer of approximately
$27.5$kW to perpendicular ion motion, mainly from ion thermal energy
($19$kW) and the $\phi\cdot j_{||}$ channel ($7$kW). This power
($27.5$kW) is likely being lost either to radial boundaries
(diagnostics for transport of perpendicular ion kinetic energy through
the boundaries have not yet been implemented) or to dissipation, but
not being returned to thermal energy in the present simulations. A
significant amount of power ($17$kW) is seen to be transferred from
ion thermal energy to perpendicular motion via the divergence of the
polarisation current term $\frac{p_i}{en_0}\nabla\cdot j_{pol}$ that
is discussed in \ref{apx:ion_energy}. The size of this term, and the
power discrepancies we observe, highlight the importance of
theoretical work to improve energetic consistency of drift-reduced
models~\cite{halpern2023} and treatment of the polarisation drift.

In summary, analysis of energy transfer channels indicates that of the
$120$kW input power approximately $73$kW goes to the divertor targets,
$9.5$kW goes to the radial boundaries, and the remaining $37.5$kW is
dissipated: $9$kW from ion parallel kinetic energy, and $27.5$kW from
ion perpendicular energy. The system has reached quasi-steady state,
but these numbers are approximate due to the large fluctuations.

These findings, and the energy flow analysis tools implemented here,
will be used to direct future improvements in the Hermes-3 model and
numerical implementation.

\section{Quantitative Validation}
\label{sec:validation}

In this section the validation methods included with the TCV-X21
dataset~\cite{tcv_x21_dataset} are applied to Hermes-3 simulation
output. After translation into a standardized format, the shared
analysis toolchain ensures as far as possible a consistent comparison
between experiment and the outputs of different simulation codes.  The
validation methodology is detailed in \cite{Oliveira_2022,Ricci_2015},
briefly summarised here.

Each row in the table below corresponds to an observable labelled $j$
from a particular diagnostic e.g. electron density from the
reciprocating Langmuir probe. This observable has $N_j$ measurements
labelled $i = 1\ldots N_j$ e.g. density at different radial locations.
The experimental measurement $i$ of an observable $j$ is $e_{j,i}$
with uncertainty (standard deviation) $\Delta e_{j,i}$. Similarly, the
simulation produces a value of $s_{j,i}$ with uncerainty $\Delta
s_{j,i}$.

The normalised difference $d_j$ between simulation and experiment for
an observable $j$ is defined in equation~\ref{eq:difference}:
\begin{equation}
  d_j = \left[\frac{1}{N_j}\sum_{i=1}^{N_j} \frac{\left(e_{j,i} - s_{j,i}\right)^2}{\Delta e_{j,i}^2 + \Delta s_{j,i}^2} \right]^{1/2} .
  \label{eq:difference}
\end{equation}
The level of agreement $R\left(d_j\right)$ smoothly varies from $0$
for perfect agreement ($d_j=0$) to $1$ for disagreement ($d_j \gg 1$).
\begin{equation}
  R\left(d_j\right) = \frac{\tanh\left[\left(d_j - 1/d_j - 1\right)/0.5\right] + 1}{2} .
\end{equation}
The sensitivity $S_j$ of an observable $j$ is a measure of the precision of the comparison
\begin{equation}
  S_j = \exp\left(-\frac{\sum_i \Delta e_{j,i} + \sum_i \Delta s_{j,i}}{\sum_i \left|e_{j,i}\right| + \sum_i \left|s_{j,i}\right|} \right) .
  \label{eq:sensitivity}
\end{equation}
$S_j$ approaches $1$ for high precision measurements, and $0$ if uncertainties are large.

The level of agreement $R\left(d_j\right)$ and sensitivities $S_j$ for
a set of observables are combined into a composite metric $\chi$ that varies between 0 (agreement) and 1 (disagreement):
\begin{equation}
  \chi = \frac{\sum_j R\left(d_j\right) H_j S_j}{\sum_j H_j S_j} . \label{eq:composite_metric}
\end{equation}
Where $H_j$ is a Hierarchy weighting that is 1 for observables that
are directly measured and simulated, smaller than 1 for derived
quantities. Table 1 of \cite{Oliveira_2022} lists $H_j$ for all
observables used here.
The denominator of equation~\ref{eq:composite_metric} is the `quality'
\begin{equation}
  Q = \sum_j H_j S_j
  \label{eq:quality}
\end{equation}
that is higher for validations using more directly computed, higher
precision observables. Table~\ref{tab:quantitative} below lists the
normalised difference $d_j$ and sensitivity $S_j$ for each
observable. These are grouped into diagnostics, with a composite
metric $\chi$ and quality $Q$. Finally an overall metric and quality
are computed by combining all diagnostics.

\begin{longtable}{llcccc}
  \caption{Quantitative validation result for each observable, in forward (+) and reversed (-) toroidal field direction. For each observable the normalized difference between experiment and simulation $d_j$ (equation~\ref{eq:difference}) and sensitivity $S_j$ (equation~\ref{eq:sensitivity}) are calculated. The color scale is limited to $0<d_j<5$; Green cells indicate good agreement ($d_j \lesssim 1$) and red cells indicate poor agreement ($d_j \gtrsim 4$). Observables are combined into a composite metric $\chi$ (equation~\ref{eq:composite_metric}) and validation quality $Q$ (equation~\ref{eq:quality}).}
  \label{tab:quantitative}
  \endfirsthead
  \endhead
\toprule
 & & \multicolumn{2}{c}{Hermes($+$)} & \multicolumn{2}{c}{Hermes($-$)} \\ 
Diagnostic & observable & $d_j$ & $S$ & $d_j$ & $S$ \\ 
\midrule
\multirow{11}{*}{\makecell{Fast\\horizontally-\\reciprocating\\probe (FHRP)\\for outboard\\midplane}}
& $n$                                      & \cellcolor[rgb]{ 0.718,  0.880,  0.459}\textcolor[rgb]{ 0.000,  0.000,  0.000}{1.01  } & \cellcolor[rgb]{ 1.000,  1.000,  1.000}\textcolor[rgb]{ 0.000,  0.000,  0.000}{0.76  } & \cellcolor[rgb]{ 0.812,  0.920,  0.520}\textcolor[rgb]{ 0.000,  0.000,  0.000}{1.44  } & \cellcolor[rgb]{ 1.000,  1.000,  1.000}\textcolor[rgb]{ 0.000,  0.000,  0.000}{0.817 }\\
& $T_e$                                    & \cellcolor[rgb]{ 0.557,  0.810,  0.405}\textcolor[rgb]{ 0.000,  0.000,  0.000}{0.388 } & \cellcolor[rgb]{ 1.000,  1.000,  1.000}\textcolor[rgb]{ 0.000,  0.000,  0.000}{0.69  } & \cellcolor[rgb]{ 0.548,  0.806,  0.404}\textcolor[rgb]{ 0.000,  0.000,  0.000}{0.331 } & \cellcolor[rgb]{ 1.000,  1.000,  1.000}\textcolor[rgb]{ 0.000,  0.000,  0.000}{0.745 }\\
& $V_{pl}$                                 & \cellcolor[rgb]{ 0.567,  0.814,  0.407}\textcolor[rgb]{ 0.000,  0.000,  0.000}{0.409 } & \cellcolor[rgb]{ 1.000,  1.000,  1.000}\textcolor[rgb]{ 0.000,  0.000,  0.000}{0.688 } & \cellcolor[rgb]{ 0.587,  0.823,  0.409}\textcolor[rgb]{ 0.000,  0.000,  0.000}{0.467 } & \cellcolor[rgb]{ 1.000,  1.000,  1.000}\textcolor[rgb]{ 0.000,  0.000,  0.000}{0.73  }\\
& $J_{sat}$                                & \cellcolor[rgb]{ 0.968,  0.986,  0.705}\textcolor[rgb]{ 0.000,  0.000,  0.000}{2.32  } & \cellcolor[rgb]{ 1.000,  1.000,  1.000}\textcolor[rgb]{ 0.000,  0.000,  0.000}{0.804 } & \cellcolor[rgb]{ 0.886,  0.952,  0.593}\textcolor[rgb]{ 0.000,  0.000,  0.000}{1.82  } & \cellcolor[rgb]{ 1.000,  1.000,  1.000}\textcolor[rgb]{ 0.000,  0.000,  0.000}{0.803 }\\
& $\sigma\left(J_{sat}\right)$             & \cellcolor[rgb]{ 0.964,  0.477,  0.286}\textcolor[rgb]{ 0.000,  0.000,  0.000}{10.3  } & \cellcolor[rgb]{ 1.000,  1.000,  1.000}\textcolor[rgb]{ 0.000,  0.000,  0.000}{0.943 } & \cellcolor[rgb]{ 0.964,  0.477,  0.286}\textcolor[rgb]{ 0.000,  0.000,  0.000}{10.7  } & \cellcolor[rgb]{ 1.000,  1.000,  1.000}\textcolor[rgb]{ 0.000,  0.000,  0.000}{0.953 }\\
& $\mathrm{skew}\left(J_{sat}\right)$      & \cellcolor[rgb]{ 0.991,  0.996,  0.737}\textcolor[rgb]{ 0.000,  0.000,  0.000}{2.45  } & \cellcolor[rgb]{ 1.000,  1.000,  1.000}\textcolor[rgb]{ 0.000,  0.000,  0.000}{0.852 } & \cellcolor[rgb]{ 0.999,  0.964,  0.689}\textcolor[rgb]{ 0.000,  0.000,  0.000}{2.75  } & \cellcolor[rgb]{ 1.000,  1.000,  1.000}\textcolor[rgb]{ 0.000,  0.000,  0.000}{0.923 }\\
& $\mathrm{kurt}\left(J_{sat}\right)$      & \cellcolor[rgb]{ 0.780,  0.907,  0.499}\textcolor[rgb]{ 0.000,  0.000,  0.000}{1.28  } & \cellcolor[rgb]{ 1.000,  1.000,  1.000}\textcolor[rgb]{ 0.000,  0.000,  0.000}{0.811 } & \cellcolor[rgb]{ 0.993,  0.748,  0.435}\textcolor[rgb]{ 0.000,  0.000,  0.000}{3.99  } & \cellcolor[rgb]{ 1.000,  1.000,  1.000}\textcolor[rgb]{ 0.000,  0.000,  0.000}{0.943 }\\
& $V_{fl}$                                 & \cellcolor[rgb]{ 0.686,  0.866,  0.439}\textcolor[rgb]{ 0.000,  0.000,  0.000}{0.879 } & \cellcolor[rgb]{ 1.000,  1.000,  1.000}\textcolor[rgb]{ 0.000,  0.000,  0.000}{0.496 } & \cellcolor[rgb]{ 0.843,  0.934,  0.540}\textcolor[rgb]{ 0.000,  0.000,  0.000}{1.57  } & \cellcolor[rgb]{ 1.000,  1.000,  1.000}\textcolor[rgb]{ 0.000,  0.000,  0.000}{0.743 }\\
& $\sigma\left(V_{fl}\right)$              & \cellcolor[rgb]{ 0.964,  0.477,  0.286}\textcolor[rgb]{ 0.000,  0.000,  0.000}{8.42  } & \cellcolor[rgb]{ 1.000,  1.000,  1.000}\textcolor[rgb]{ 0.000,  0.000,  0.000}{0.953 } & \cellcolor[rgb]{ 0.964,  0.477,  0.286}\textcolor[rgb]{ 0.000,  0.000,  0.000}{5.52  } & \cellcolor[rgb]{ 1.000,  1.000,  1.000}\textcolor[rgb]{ 0.000,  0.000,  0.000}{0.934 }\\
& $M_\parallel$                            & \cellcolor[rgb]{ 1.000,  0.993,  0.737}\textcolor[rgb]{ 0.000,  0.000,  0.000}{2.56  } & \cellcolor[rgb]{ 1.000,  1.000,  1.000}\textcolor[rgb]{ 0.000,  0.000,  0.000}{0.816 } & \cellcolor[rgb]{ 0.994,  0.778,  0.461}\textcolor[rgb]{ 0.000,  0.000,  0.000}{3.86  } & \cellcolor[rgb]{ 1.000,  1.000,  1.000}\textcolor[rgb]{ 0.000,  0.000,  0.000}{0.847 }\\
\cline{2-6}
& $\left(\chi; Q\right)$\textsubscript{FHRP} & \multicolumn{2}{c}{$ \textbf{(0.55; \ 3.77)} $ } & \multicolumn{2}{c}{$ \textbf{(0.67; \ 4.08)} $ }\\
\midrule
\multirow{3}{*}{\makecell{Thomson scattering\\(TS) for divertor\\entrance}}
& $n$                                      & \cellcolor[rgb]{ 0.757,  0.897,  0.484}\textcolor[rgb]{ 0.000,  0.000,  0.000}{1.19  } & \cellcolor[rgb]{ 1.000,  1.000,  1.000}\textcolor[rgb]{ 0.000,  0.000,  0.000}{0.842 } & \cellcolor[rgb]{ 0.835,  0.930,  0.535}\textcolor[rgb]{ 0.000,  0.000,  0.000}{1.55  } & \cellcolor[rgb]{ 1.000,  1.000,  1.000}\textcolor[rgb]{ 0.000,  0.000,  0.000}{0.854 }\\
& $T_e$                                    & \cellcolor[rgb]{ 0.577,  0.819,  0.408}\textcolor[rgb]{ 0.000,  0.000,  0.000}{0.43  } & \cellcolor[rgb]{ 1.000,  1.000,  1.000}\textcolor[rgb]{ 0.000,  0.000,  0.000}{0.858 } & \cellcolor[rgb]{ 0.678,  0.863,  0.433}\textcolor[rgb]{ 0.000,  0.000,  0.000}{0.83  } & \cellcolor[rgb]{ 1.000,  1.000,  1.000}\textcolor[rgb]{ 0.000,  0.000,  0.000}{0.872 }\\
\cline{2-6}
& $\left(\chi; Q\right)$\textsubscript{TS} & \multicolumn{2}{c}{$ \textbf{(0.035; \ 0.85)} $ } & \multicolumn{2}{c}{$ \textbf{(0.2; \ 0.863)} $ }\\
\midrule
\multirow{11}{*}{\makecell{Reciprocating\\divertor probe\\array (RDPA)\\for divertor\\volume}}
& $n$                                      & \cellcolor[rgb]{ 0.998,  0.940,  0.649}\textcolor[rgb]{ 0.000,  0.000,  0.000}{2.93  } & \cellcolor[rgb]{ 1.000,  1.000,  1.000}\textcolor[rgb]{ 0.000,  0.000,  0.000}{0.854 } & \cellcolor[rgb]{ 0.964,  0.477,  0.286}\textcolor[rgb]{ 0.000,  0.000,  0.000}{5.07  } & \cellcolor[rgb]{ 1.000,  1.000,  1.000}\textcolor[rgb]{ 0.000,  0.000,  0.000}{0.845 }\\
& $T_e$                                    & \cellcolor[rgb]{ 0.933,  0.972,  0.657}\textcolor[rgb]{ 0.000,  0.000,  0.000}{2.11  } & \cellcolor[rgb]{ 1.000,  1.000,  1.000}\textcolor[rgb]{ 0.000,  0.000,  0.000}{0.889 } & \cellcolor[rgb]{ 0.927,  0.969,  0.649}\textcolor[rgb]{ 0.000,  0.000,  0.000}{2.06  } & \cellcolor[rgb]{ 1.000,  1.000,  1.000}\textcolor[rgb]{ 0.000,  0.000,  0.000}{0.889 }\\
& $V_{pl}$                                 & \cellcolor[rgb]{ 0.820,  0.924,  0.525}\textcolor[rgb]{ 0.000,  0.000,  0.000}{1.47  } & \cellcolor[rgb]{ 1.000,  1.000,  1.000}\textcolor[rgb]{ 0.000,  0.000,  0.000}{0.877 } & \cellcolor[rgb]{ 0.773,  0.903,  0.494}\textcolor[rgb]{ 0.000,  0.000,  0.000}{1.26  } & \cellcolor[rgb]{ 1.000,  1.000,  1.000}\textcolor[rgb]{ 0.000,  0.000,  0.000}{0.887 }\\
& $J_{sat}$                                & \cellcolor[rgb]{ 0.996,  0.863,  0.532}\textcolor[rgb]{ 0.000,  0.000,  0.000}{3.47  } & \cellcolor[rgb]{ 1.000,  1.000,  1.000}\textcolor[rgb]{ 0.000,  0.000,  0.000}{0.862 } & \cellcolor[rgb]{ 0.927,  0.969,  0.649}\textcolor[rgb]{ 0.000,  0.000,  0.000}{2.05  } & \cellcolor[rgb]{ 1.000,  1.000,  1.000}\textcolor[rgb]{ 0.000,  0.000,  0.000}{0.849 }\\
& $\sigma\left(J_{sat}\right)$             & \cellcolor[rgb]{ 0.997,  0.898,  0.577}\textcolor[rgb]{ 0.000,  0.000,  0.000}{3.27  } & \cellcolor[rgb]{ 1.000,  1.000,  1.000}\textcolor[rgb]{ 0.000,  0.000,  0.000}{0.865 } & \cellcolor[rgb]{ 0.964,  0.477,  0.286}\textcolor[rgb]{ 0.000,  0.000,  0.000}{8.51  } & \cellcolor[rgb]{ 1.000,  1.000,  1.000}\textcolor[rgb]{ 0.000,  0.000,  0.000}{0.851 }\\
& $\mathrm{skew}\left(J_{sat}\right)$      & \cellcolor[rgb]{ 0.997,  0.898,  0.577}\textcolor[rgb]{ 0.000,  0.000,  0.000}{3.27  } & \cellcolor[rgb]{ 1.000,  1.000,  1.000}\textcolor[rgb]{ 0.000,  0.000,  0.000}{0.812 } & \cellcolor[rgb]{ 0.974,  0.989,  0.713}\textcolor[rgb]{ 0.000,  0.000,  0.000}{2.34  } & \cellcolor[rgb]{ 1.000,  1.000,  1.000}\textcolor[rgb]{ 0.000,  0.000,  0.000}{0.789 }\\
& $\mathrm{kurt}\left(J_{sat}\right)$      & \cellcolor[rgb]{ 0.998,  0.950,  0.665}\textcolor[rgb]{ 0.000,  0.000,  0.000}{2.86  } & \cellcolor[rgb]{ 1.000,  1.000,  1.000}\textcolor[rgb]{ 0.000,  0.000,  0.000}{0.9   } & \cellcolor[rgb]{ 0.965,  0.487,  0.290}\textcolor[rgb]{ 0.000,  0.000,  0.000}{5.0   } & \cellcolor[rgb]{ 1.000,  1.000,  1.000}\textcolor[rgb]{ 0.000,  0.000,  0.000}{0.921 }\\
& $V_{fl}$                                 & \cellcolor[rgb]{ 0.944,  0.977,  0.673}\textcolor[rgb]{ 0.000,  0.000,  0.000}{2.17  } & \cellcolor[rgb]{ 1.000,  1.000,  1.000}\textcolor[rgb]{ 0.000,  0.000,  0.000}{0.657 } & \cellcolor[rgb]{ 0.921,  0.967,  0.641}\textcolor[rgb]{ 0.000,  0.000,  0.000}{2.04  } & \cellcolor[rgb]{ 1.000,  1.000,  1.000}\textcolor[rgb]{ 0.000,  0.000,  0.000}{0.77  }\\
& $\sigma\left(V_{fl}\right)$              & \cellcolor[rgb]{ 0.964,  0.477,  0.286}\textcolor[rgb]{ 0.000,  0.000,  0.000}{41.6  } & \cellcolor[rgb]{ 1.000,  1.000,  1.000}\textcolor[rgb]{ 0.000,  0.000,  0.000}{0.928 } & \cellcolor[rgb]{ 0.964,  0.477,  0.286}\textcolor[rgb]{ 0.000,  0.000,  0.000}{11.0  } & \cellcolor[rgb]{ 1.000,  1.000,  1.000}\textcolor[rgb]{ 0.000,  0.000,  0.000}{0.886 }\\
& $M_\parallel$                            & \cellcolor[rgb]{ 0.964,  0.477,  0.286}\textcolor[rgb]{ 0.000,  0.000,  0.000}{6.54  } & \cellcolor[rgb]{ 1.000,  1.000,  1.000}\textcolor[rgb]{ 0.000,  0.000,  0.000}{0.862 } & \cellcolor[rgb]{ 0.964,  0.477,  0.286}\textcolor[rgb]{ 0.000,  0.000,  0.000}{11.4  } & \cellcolor[rgb]{ 1.000,  1.000,  1.000}\textcolor[rgb]{ 0.000,  0.000,  0.000}{0.862 }\\
\cline{2-6}
& $\left(\chi; Q\right)$\textsubscript{RDPA} & \multicolumn{2}{c}{$ \textbf{(0.91; \ 4.11)} $ } & \multicolumn{2}{c}{$ \textbf{(0.87; \ 4.13)} $ }\\
\midrule
\multirow{2}{*}{\makecell{Infrared camera (IR)\\for low-field-side target}}
& $q_\parallel$                            & \cellcolor[rgb]{ 0.964,  0.477,  0.286}\textcolor[rgb]{ 0.000,  0.000,  0.000}{5.22  } & \cellcolor[rgb]{ 1.000,  1.000,  1.000}\textcolor[rgb]{ 0.000,  0.000,  0.000}{0.808 } & \cellcolor[rgb]{ 0.964,  0.477,  0.286}\textcolor[rgb]{ 0.000,  0.000,  0.000}{5.67  } & \cellcolor[rgb]{ 1.000,  1.000,  1.000}\textcolor[rgb]{ 0.000,  0.000,  0.000}{0.839 }\\
\cline{2-6}
& $\left(\chi; Q\right)$\textsubscript{LFS-IR} & \multicolumn{2}{c}{$ \textbf{(1.0; \ 0.404)} $ } & \multicolumn{2}{c}{$ \textbf{(1.0; \ 0.42)} $ }\\
\midrule
\multirow{12}{*}{\makecell{Wall Langmuir\\probes for\\low-field-side\\target}}
& $n$                                      & \cellcolor[rgb]{ 1.000,  0.993,  0.737}\textcolor[rgb]{ 0.000,  0.000,  0.000}{2.56  } & \cellcolor[rgb]{ 1.000,  1.000,  1.000}\textcolor[rgb]{ 0.000,  0.000,  0.000}{0.81  } & \cellcolor[rgb]{ 0.820,  0.924,  0.525}\textcolor[rgb]{ 0.000,  0.000,  0.000}{1.48  } & \cellcolor[rgb]{ 1.000,  1.000,  1.000}\textcolor[rgb]{ 0.000,  0.000,  0.000}{0.786 }\\
& $T_e$                                    & \cellcolor[rgb]{ 0.939,  0.974,  0.665}\textcolor[rgb]{ 0.000,  0.000,  0.000}{2.12  } & \cellcolor[rgb]{ 1.000,  1.000,  1.000}\textcolor[rgb]{ 0.000,  0.000,  0.000}{0.882 } & \cellcolor[rgb]{ 0.892,  0.954,  0.601}\textcolor[rgb]{ 0.000,  0.000,  0.000}{1.85  } & \cellcolor[rgb]{ 1.000,  1.000,  1.000}\textcolor[rgb]{ 0.000,  0.000,  0.000}{0.865 }\\
& $V_{pl}$                                 & \cellcolor[rgb]{ 0.999,  0.969,  0.697}\textcolor[rgb]{ 0.000,  0.000,  0.000}{2.73  } & \cellcolor[rgb]{ 1.000,  1.000,  1.000}\textcolor[rgb]{ 0.000,  0.000,  0.000}{0.88  } & \cellcolor[rgb]{ 0.851,  0.937,  0.545}\textcolor[rgb]{ 0.000,  0.000,  0.000}{1.61  } & \cellcolor[rgb]{ 1.000,  1.000,  1.000}\textcolor[rgb]{ 0.000,  0.000,  0.000}{0.88  }\\
& $J_{sat}$                                & \cellcolor[rgb]{ 0.996,  0.888,  0.561}\textcolor[rgb]{ 0.000,  0.000,  0.000}{3.31  } & \cellcolor[rgb]{ 1.000,  1.000,  1.000}\textcolor[rgb]{ 0.000,  0.000,  0.000}{0.848 } & \cellcolor[rgb]{ 0.927,  0.969,  0.649}\textcolor[rgb]{ 0.000,  0.000,  0.000}{2.05  } & \cellcolor[rgb]{ 1.000,  1.000,  1.000}\textcolor[rgb]{ 0.000,  0.000,  0.000}{0.826 }\\
& $\sigma\left(J_{sat}\right)$             & \cellcolor[rgb]{ 0.980,  0.597,  0.341}\textcolor[rgb]{ 0.000,  0.000,  0.000}{4.6   } & \cellcolor[rgb]{ 1.000,  1.000,  1.000}\textcolor[rgb]{ 0.000,  0.000,  0.000}{0.85  } & \cellcolor[rgb]{ 0.992,  0.694,  0.390}\textcolor[rgb]{ 0.000,  0.000,  0.000}{4.25  } & \cellcolor[rgb]{ 1.000,  1.000,  1.000}\textcolor[rgb]{ 0.000,  0.000,  0.000}{0.865 }\\
& $\mathrm{skew}\left(J_{sat}\right)$      & \cellcolor[rgb]{ 0.998,  0.936,  0.641}\textcolor[rgb]{ 0.000,  0.000,  0.000}{2.97  } & \cellcolor[rgb]{ 1.000,  1.000,  1.000}\textcolor[rgb]{ 0.000,  0.000,  0.000}{0.833 } & \cellcolor[rgb]{ 0.964,  0.477,  0.286}\textcolor[rgb]{ 0.000,  0.000,  0.000}{42.8  } & \cellcolor[rgb]{ 1.000,  1.000,  1.000}\textcolor[rgb]{ 0.000,  0.000,  0.000}{0.956 }\\
& $\mathrm{kurt}\left(J_{sat}\right)$      & \cellcolor[rgb]{ 0.989,  0.657,  0.369}\textcolor[rgb]{ 0.000,  0.000,  0.000}{4.39  } & \cellcolor[rgb]{ 1.000,  1.000,  1.000}\textcolor[rgb]{ 0.000,  0.000,  0.000}{0.907 } & \cellcolor[rgb]{ 0.964,  0.477,  0.286}\textcolor[rgb]{ 0.000,  0.000,  0.000}{40.8  } & \cellcolor[rgb]{ 1.000,  1.000,  1.000}\textcolor[rgb]{ 0.000,  0.000,  0.000}{0.98  }\\
& $J_\parallel$                            & \cellcolor[rgb]{ 0.956,  0.982,  0.689}\textcolor[rgb]{ 0.000,  0.000,  0.000}{2.23  } & \cellcolor[rgb]{ 1.000,  1.000,  1.000}\textcolor[rgb]{ 0.000,  0.000,  0.000}{0.619 } & \cellcolor[rgb]{ 0.741,  0.890,  0.474}\textcolor[rgb]{ 0.000,  0.000,  0.000}{1.12  } & \cellcolor[rgb]{ 1.000,  1.000,  1.000}\textcolor[rgb]{ 0.000,  0.000,  0.000}{0.668 }\\
& $\sigma\left(J_\parallel\right)$         & \cellcolor[rgb]{ 0.993,  0.725,  0.416}\textcolor[rgb]{ 0.000,  0.000,  0.000}{4.09  } & \cellcolor[rgb]{ 1.000,  1.000,  1.000}\textcolor[rgb]{ 0.000,  0.000,  0.000}{0.857 } & \cellcolor[rgb]{ 0.996,  0.883,  0.553}\textcolor[rgb]{ 0.000,  0.000,  0.000}{3.36  } & \cellcolor[rgb]{ 1.000,  1.000,  1.000}\textcolor[rgb]{ 0.000,  0.000,  0.000}{0.852 }\\
& $V_{fl}$                                 & \cellcolor[rgb]{ 0.678,  0.863,  0.433}\textcolor[rgb]{ 0.000,  0.000,  0.000}{0.841 } & \cellcolor[rgb]{ 1.000,  1.000,  1.000}\textcolor[rgb]{ 0.000,  0.000,  0.000}{0.614 } & \cellcolor[rgb]{ 0.904,  0.959,  0.617}\textcolor[rgb]{ 0.000,  0.000,  0.000}{1.93  } & \cellcolor[rgb]{ 1.000,  1.000,  1.000}\textcolor[rgb]{ 0.000,  0.000,  0.000}{0.707 }\\
& $\sigma\left(V_{fl}\right)$              & \cellcolor[rgb]{ 0.964,  0.477,  0.286}\textcolor[rgb]{ 0.000,  0.000,  0.000}{6.87  } & \cellcolor[rgb]{ 1.000,  1.000,  1.000}\textcolor[rgb]{ 0.000,  0.000,  0.000}{0.91  } & \cellcolor[rgb]{ 0.964,  0.477,  0.286}\textcolor[rgb]{ 0.000,  0.000,  0.000}{8.83  } & \cellcolor[rgb]{ 1.000,  1.000,  1.000}\textcolor[rgb]{ 0.000,  0.000,  0.000}{0.925 }\\
\cline{2-6}
& $\left(\chi; Q\right)$\textsubscript{LFS-LP} & \multicolumn{2}{c}{$ \textbf{(0.93; \ 5.24)} $ } & \multicolumn{2}{c}{$ \textbf{(0.75; \ 5.41)} $ }\\
\midrule
\multirow{12}{*}{\makecell{Wall Langmuir\\probes for\\high-field-side\\target}}
& $n$                                      & \cellcolor[rgb]{ 0.995,  0.848,  0.519}\textcolor[rgb]{ 0.000,  0.000,  0.000}{3.55  } & \cellcolor[rgb]{ 1.000,  1.000,  1.000}\textcolor[rgb]{ 0.000,  0.000,  0.000}{0.842 } & \cellcolor[rgb]{ 0.827,  0.927,  0.530}\textcolor[rgb]{ 0.000,  0.000,  0.000}{1.49  } & \cellcolor[rgb]{ 1.000,  1.000,  1.000}\textcolor[rgb]{ 0.000,  0.000,  0.000}{0.809 }\\
& $T_e$                                    & \cellcolor[rgb]{ 0.995,  0.848,  0.519}\textcolor[rgb]{ 0.000,  0.000,  0.000}{3.52  } & \cellcolor[rgb]{ 1.000,  1.000,  1.000}\textcolor[rgb]{ 0.000,  0.000,  0.000}{0.906 } & \cellcolor[rgb]{ 0.725,  0.883,  0.464}\textcolor[rgb]{ 0.000,  0.000,  0.000}{1.03  } & \cellcolor[rgb]{ 1.000,  1.000,  1.000}\textcolor[rgb]{ 0.000,  0.000,  0.000}{0.88  }\\
& $V_{pl}$                                 & \cellcolor[rgb]{ 0.921,  0.967,  0.641}\textcolor[rgb]{ 0.000,  0.000,  0.000}{2.04  } & \cellcolor[rgb]{ 1.000,  1.000,  1.000}\textcolor[rgb]{ 0.000,  0.000,  0.000}{0.904 } & \cellcolor[rgb]{ 0.733,  0.887,  0.469}\textcolor[rgb]{ 0.000,  0.000,  0.000}{1.09  } & \cellcolor[rgb]{ 1.000,  1.000,  1.000}\textcolor[rgb]{ 0.000,  0.000,  0.000}{0.882 }\\
& $J_{sat}$                                & \cellcolor[rgb]{ 0.999,  0.983,  0.721}\textcolor[rgb]{ 0.000,  0.000,  0.000}{2.62  } & \cellcolor[rgb]{ 1.000,  1.000,  1.000}\textcolor[rgb]{ 0.000,  0.000,  0.000}{0.827 } & \cellcolor[rgb]{ 0.997,  0.999,  0.745}\textcolor[rgb]{ 0.000,  0.000,  0.000}{2.46  } & \cellcolor[rgb]{ 1.000,  1.000,  1.000}\textcolor[rgb]{ 0.000,  0.000,  0.000}{0.808 }\\
& $\sigma\left(J_{sat}\right)$             & \cellcolor[rgb]{ 0.991,  0.677,  0.378}\textcolor[rgb]{ 0.000,  0.000,  0.000}{4.3   } & \cellcolor[rgb]{ 1.000,  1.000,  1.000}\textcolor[rgb]{ 0.000,  0.000,  0.000}{0.821 } & \cellcolor[rgb]{ 0.993,  0.732,  0.422}\textcolor[rgb]{ 0.000,  0.000,  0.000}{4.06  } & \cellcolor[rgb]{ 1.000,  1.000,  1.000}\textcolor[rgb]{ 0.000,  0.000,  0.000}{0.881 }\\
& $\mathrm{skew}\left(J_{sat}\right)$      & \cellcolor[rgb]{ 0.990,  0.667,  0.373}\textcolor[rgb]{ 0.000,  0.000,  0.000}{4.36  } & \cellcolor[rgb]{ 1.000,  1.000,  1.000}\textcolor[rgb]{ 0.000,  0.000,  0.000}{0.814 } & \cellcolor[rgb]{ 0.964,  0.477,  0.286}\textcolor[rgb]{ 0.000,  0.000,  0.000}{11.8  } & \cellcolor[rgb]{ 1.000,  1.000,  1.000}\textcolor[rgb]{ 0.000,  0.000,  0.000}{0.888 }\\
& $\mathrm{kurt}\left(J_{sat}\right)$      & \cellcolor[rgb]{ 0.964,  0.477,  0.286}\textcolor[rgb]{ 0.000,  0.000,  0.000}{8.66  } & \cellcolor[rgb]{ 1.000,  1.000,  1.000}\textcolor[rgb]{ 0.000,  0.000,  0.000}{0.94  } & \cellcolor[rgb]{ 0.964,  0.477,  0.286}\textcolor[rgb]{ 0.000,  0.000,  0.000}{19.2  } & \cellcolor[rgb]{ 1.000,  1.000,  1.000}\textcolor[rgb]{ 0.000,  0.000,  0.000}{0.973 }\\
& $J_\parallel$                            & \cellcolor[rgb]{ 0.944,  0.977,  0.673}\textcolor[rgb]{ 0.000,  0.000,  0.000}{2.17  } & \cellcolor[rgb]{ 1.000,  1.000,  1.000}\textcolor[rgb]{ 0.000,  0.000,  0.000}{0.676 } & \cellcolor[rgb]{ 0.857,  0.940,  0.553}\textcolor[rgb]{ 0.000,  0.000,  0.000}{1.65  } & \cellcolor[rgb]{ 1.000,  1.000,  1.000}\textcolor[rgb]{ 0.000,  0.000,  0.000}{0.745 }\\
& $\sigma\left(J_\parallel\right)$         & \cellcolor[rgb]{ 0.995,  0.832,  0.506}\textcolor[rgb]{ 0.000,  0.000,  0.000}{3.6   } & \cellcolor[rgb]{ 1.000,  1.000,  1.000}\textcolor[rgb]{ 0.000,  0.000,  0.000}{0.88  } & \cellcolor[rgb]{ 0.997,  0.917,  0.609}\textcolor[rgb]{ 0.000,  0.000,  0.000}{3.13  } & \cellcolor[rgb]{ 1.000,  1.000,  1.000}\textcolor[rgb]{ 0.000,  0.000,  0.000}{0.882 }\\
& $V_{fl}$                                 & \cellcolor[rgb]{ 1.000,  0.998,  0.745}\textcolor[rgb]{ 0.000,  0.000,  0.000}{2.51  } & \cellcolor[rgb]{ 1.000,  1.000,  1.000}\textcolor[rgb]{ 0.000,  0.000,  0.000}{0.618 } & \cellcolor[rgb]{ 0.617,  0.836,  0.412}\textcolor[rgb]{ 0.000,  0.000,  0.000}{0.571 } & \cellcolor[rgb]{ 1.000,  1.000,  1.000}\textcolor[rgb]{ 0.000,  0.000,  0.000}{0.719 }\\
& $\sigma\left(V_{fl}\right)$              & \cellcolor[rgb]{ 0.997,  0.907,  0.593}\textcolor[rgb]{ 0.000,  0.000,  0.000}{3.19  } & \cellcolor[rgb]{ 1.000,  1.000,  1.000}\textcolor[rgb]{ 0.000,  0.000,  0.000}{0.898 } & \cellcolor[rgb]{ 0.964,  0.477,  0.286}\textcolor[rgb]{ 0.000,  0.000,  0.000}{15.6  } & \cellcolor[rgb]{ 1.000,  1.000,  1.000}\textcolor[rgb]{ 0.000,  0.000,  0.000}{0.913 }\\
\cline{2-6}
& $\left(\chi; Q\right)$\textsubscript{HFS-LP} & \multicolumn{2}{c}{$ \textbf{(0.98; \ 5.34)} $ } & \multicolumn{2}{c}{$ \textbf{(0.67; \ 5.5)} $ }\\
\midrule
Overall
& $\chi$; $Q$                              & \multicolumn{2}{c}{$ \textbf{(0.83; \ 19.7)} $ } & \multicolumn{2}{c}{$ \textbf{(0.72; \ 20.4)} $ }\\
\bottomrule
\end{longtable}

These results compare favorably to the results of other simulation
codes in the TCV-X21 validation study~\cite{Oliveira_2022} and
quantify observations made in earlier sections. For example, both
high- and low-field target Langmuir probe measurements are better
predicted in reverse field than forward field.

A significant difference that has not been discussed in earlier
sections is in the parallel flow Mach number $M_{||}$, in both the
Fast Horizontally-Reciprocating Probe (FHRP) and the Reciprocating
Divertor Probe Array (RDPA) diagnostics. This is consistent with the
particle source (ionization) being different in simulation and
experiment: In simulation the particle source in the core leads to
strong flows into the divertor, whereas in experiment the particle
source is close to the targets. The codes used in the original
study~\cite{Oliveira_2022}, that also did not include neutrals,
observed comparable differences.

It can also be seen in table~\ref{tab:quantitative}
that in general the mean profiles (e.g. density, $J_{sat}$, $J_{||}$)
are better predicted than higher moments of the fluctuations
(e.g. $\sigma\left(J_{sat}\right)$), in common with other codes in the
study~\cite{Oliveira_2022}.  It is intriguing that the transport could
be better captured than the characteristics of the turbulence that is
in large part responsible for the transport.

\section{Conclusions}
\label{sec:conclusions}

3D turbulence simulations have been performed using
Hermes-3~\cite{hermes-3-paper}, of the TCV-X21 reference
cases~\cite{Oliveira_2022}: Two lower single-null sheath-limited TCV
L-mode scenarios with opposite toroidal magnetic field directions,
designated forward (favorable $\nabla B$) and reverse (unfavorable
$\nabla B$). Electrostatic drift-reduced fluid simulations have been
performed in a field-aligned coordinate system, evolving electron
density, ion and electron temperatures, electron and ion parallel
momentum, and vorticity with Oberbeck-Boussinesq approximation.  The
equations are derived from~\cite{simakov-2003, catto-2004,
  simakov-2004} with modifications described in
\ref{apx:modifications}.  The simulations are flux-driven, evolving
total plasma quantities without separation into profiles and
fluctuations, and are driven by fixed input power and particle fluxes
inferred from experiment~\cite{Oliveira_2022}, $120$kW and $3\times
10^{21}$ particles per second.

The simulation results reproduce many qualitative and quantitative
features of experiment, without any tuning of simulation input
parameters. Agreement is particularly good in the near SOL, where the
expected ion sound gyro-radius spatial scales are well resolved. The
computational expense of the simulations shown here was approximately
6000 core-hours per ms, running on 64 cores, with several ms of
simulation time required to relax profiles to quasi-steady state.
These simulations demonstrate the ability of Hermes-3 to robustly
perform long simulations of medium-sized tokamak (MST) edge turbulence
at modest computational expense. Detailed analysis of the energy
balance in the simulations (section~\ref{sec:heat_fluxes}) points to
the need for further improvements in the handling of perpendicular
flow energy that is related to the vorticity equation and polarisation
current. Ongoing development aims to ensure that dissipated
perpendicular flow energy goes into ion heating, making use of
improved polarisation current formulation~\cite{halpern2023}.

Differences between simulation and experiment point to the role of
neutrals in cooling the divertor targets, replacing particle flux from
the core with ionisation in the divertor legs.  One of the conclusions
of the original TCV-X21 study~\cite{Oliveira_2022} was that despite
the choice of a sheath-limited experimental regime, plasma-neutral
interactions were likely influencing the turbulence in the
divertor. Simulation of the TCV-X21 cases with
SOLPS-ITER~\cite{Wang_2024}, including kinetic neutrals, found that
$\sim 65$\% of the total ionization occurs in the SOL and
divertor. Those results support findings that neutrals are important
even in these sheath-limited plasma scenarios. The SOLPS-ITER
simulations were performed without drifts, so cannot explain
differences between forward and reversed field configurations. The
experimental data~\cite{Wang_2024} shows higher Balmer line radiation
(Divertor spectroscopy system) and higher neutral gas pressure
(barotron measurements) in forward field, indicating stronger
plasma-neutral interaction. Including these interactions to improve
model fidelity is the subject of ongoing work.

\section*{Acknowledgements}

Prepared by LLNL under Contract DE-AC52-07NA27344.  This work made use
of the version 1.0 of the TCV-X21 experimental
datasets~\cite{tcv_x21_dataset} provided by the TCV team and publicly
released under the CC-BY 4.0 license. B.D. thanks Tom Body for many
helpful discussions. The results shown here were produced with
Hermes-3 version \texttt{05d4d98} linked to BOUT++ version
\texttt{a776c8c}. Hermes-3 (LLNL-CODE-845139) is available on
Github~\cite{dudson:hermes3} under the GPL-3 license.
LLNL-JRNL-2005230.

\section*{Government Use License Notice}

This manuscript has been authored by Lawrence Livermore National
Security, LLC under Contract No. DE-AC52-07NA27344 with the
US. Department of Energy. The United States Government retains, and
the publisher, by accepting the article for publication, acknowledges
that the United States Government retains a non-exclusive, paid-up,
irrevocable, world-wide license to publish or reproduce the published
form of this manuscript, or allow others to do so, for United States
Government purposes.

\section*{References}
\bibliography{bibliography}
\bibliographystyle{unsrt}

\appendix
\newpage

\section{Modifications to drift-reduced fluid equations}
\label{apx:modifications}

The drift-reduced fluid equations implemented in Hermes-3 have been
modified from those derived in \cite{simakov-2003, catto-2004,
simakov-2004} so that they (a) make the electrostatic approximation;
(b) Have a conserved energy that includes
the perpendicular ion flow, bounding the energy in plasma flows; (c)
Make the Oberbeck-Boussinesq approximation, simplifying the
calculation of potential $\phi$ from vorticity $\omega$
(equation~\ref{eq:vorticity_definition}).

\subsection{Vorticity equation}
\label{apx:vorticity}

The form of vorticity used here (equation~\ref{eq:vorticity}) is
adapted from~\cite{simakov-2003, catto-2004, simakov-2004}. Using SI
units and the notation used in section~\ref{sec:model} the vorticity
equation is:
\begin{numparts}
\begin{eqnarray}
\frac{\partial\omega}{\partial t} &=& \nabla\cdot\mathbf{J}_{||} + \nabla\times\frac{\mathbf{b}}{B}\cdot\nabla\left(p_i + p_e\right) \\
&&-\nabla\cdot\left[\nabla_\perp \left(\frac{m_i}{2B^2}\mathbf{v}_E \cdot\nabla p_i\right) + \frac{\omega}{2}\mathbf{v}_E\right] \label{eq:scvort_b}\\
&&-\nabla\cdot\left[\left(\frac{m_in}{2B^2}\nabla_\perp^2\phi\right)\left(\mathbf{v}_E + \frac{\mathbf{b}\times\nabla p_i}{enB} \right) \right] \label{eq:scvort_c}\\
&&-\nabla\cdot\left[\left(\frac{m_i}{2B^2}\mathbf{v}_E\cdot\nabla n\right)\nabla_\perp\phi\right] \label{eq:scvort_d} \\
&&+\nabla\cdot\left[\frac{\mathbf{b}}{B}\times\left(\mathbf{\kappa}\pi_{ci}-\frac{1}{3}\nabla\pi_{ci} - \mathbf{S}^M\right)\right]
\end{eqnarray}
\end{numparts}
In the Hermes-3 implementation the Oberbeck-Boussinesq approximation
is made, so that in the polarisation current terms $n\rightarrow n_0$
(a constant), $\nabla n\rightarrow 0$, and so
equation~\ref{eq:scvort_d} is removed.

The other significant change is in equation~\ref{eq:scvort_b},
motivated by conservation of energy. Energy transfer channels
(see~\ref{apx:energy}) are calculated by multiplying the vorticity
equation by $\left(\phi + \frac{p_i}{en_0}\right)$. The first term in
equation~\ref{eq:scvort_b} becomes:
\begin{numparts}
\begin{eqnarray}
\left(\phi + \frac{p_i}{en_0}\right) && \nabla\cdot\left[\nabla_\perp \left(\frac{m_i}{2B^2}\mathbf{v}_E \cdot\nabla p_i\right)\right] \nonumber \\
&&= \nabla\cdot\left[\frac{m_i}{2B^2}\left(\phi + \frac{p_i}{en_0}\right)\nabla_\perp \left(\mathbf{v}_E \cdot\nabla p_i\right)\right] \label{eq:scvort_b_a}\\
&& - \nabla_\perp\left[ \frac{m_i}{2B^2} \left(\mathbf{v}_E \cdot\nabla p_i\right)\nabla\left(\phi + \frac{p_i}{en_0}\right) \right] \label{eq:scvort_b_b}\\
&& + \left(\mathbf{v}_E \cdot\nabla p_i\right)\frac{\omega}{2} \label{eq:scvort_b_c} \\
&& + \left(\phi + \frac{p_i}{en_0}\right)\nabla\cdot\left[\left(\mathbf{v}_E \cdot\nabla p_i\right)\nabla_\perp \left(\frac{m_i}{2B^2}\right)\right] \label{eq:scvort_b_d}
\end{eqnarray}
\end{numparts}

The terms~\ref{eq:scvort_b_a} and~\ref{eq:scvort_b_b} are divergences,
and so integrate over the domain to leave only boundary fluxes.
Term~\ref{eq:scvort_b_c} cancels the energy transfer term coming from
the second term in equation~\ref{eq:scvort_b}. The final term, \ref{eq:scvort_b_d}, is a spurious source of energy if there are gradients in the magnetic field strength $B$.

The polarisation current terms, including equation~\ref{eq:scvort_b},
are derived in \cite{simakov-2003, simakov-2004} in a straight
magnetic field (i.e. constant $B$). Consistent with that
approximation, we may move the factor of $B^2$ through the spatial
derivative in equation~\ref{eq:scvort_b} such that \ref{eq:scvort_b_d}
is removed and energy conservation restored. This leads to the form of
the vorticity equation implemented in Hermes-3
(equation~\ref{eq:vorticity}).

\subsection{Ion energy}
\label{apx:ion_energy}

The term $\frac{p_i}{en_0}\nabla\cdot\left(\mathbf{J}_{||}
+ \mathbf{J}_d\right)$ appearing in the ion thermal energy
(equation~\ref{eq:energy_pi}) is compression due to the ion
polarisation velocity and viscous terms. This can be seen from
quasineutrality: $\nabla\cdot\left(\mathbf{J}_{||}
+ \mathbf{J}_d\right) = -\nabla\cdot\left(\mathbf{J}_{pol}
+ \mathbf{J}_{ci}\right)$. Taking the Oberbeck-Boussinesq
approximation $\mathbf{J}_{pol} = en_0\mathbf{v}_{pol}$, and so
$\frac{p_i}{en_0}\nabla\cdot\left(\mathbf{J}_{||}
+ \mathbf{J}_d\right)\simeq -p_i\nabla\cdot\mathbf{v}_{pol}$.
This term therefore adds to similar compression terms for
the $E\times B$ and parallel flow.

This ion pressure term can be derived from the vorticity equation
using energy conservation considerations (\ref{apx:energy}), or from
first principles~\cite{madsen2016a} as implemented in
e.g. HESEL~\cite{Nielsen_2017}. It does not appear in the equations of
Simakov \& Catto~\cite{simakov-2003, catto-2004, simakov-2004} because
those equations do not include the perpendicular flow in the conserved
energy of the system, and polarisation drift is formally smaller than
$E\times B$ drift in the drift ordering.

\section{Energy conservation}
\label{apx:energy}

The set of equations given in section~\ref{sec:model} has a conserved energy
\begin{equation}
E = \int \frac{1}{2}n_0m_i\left|\frac{\nabla_\perp\phi}{B} + \frac{\nabla_\perp p_i}{e n_0 B}\right|^2 + \frac{3}{2}\left(p_e + p_i\right) + \frac{1}{2} m_inv_{||i}^2 + \frac{1}{2} m_env_{||e}^2 dV 
\end{equation}
where the integral is over the domain volume $V$. This energy includes
both the perpendicular and parallel flow kinetic energy.

Using the methodology of B.Scott~\cite{Scott_2007}, energy
conservation can be evaluated by multiplying the vorticity equation by
$\left(\phi + \frac{p_i}{en_0}\right)$, the ion momentum equation by
$v_{||i}$ and the electron momentum equation by $v_{||e}$ and
rearranging.  Terms that can be written as divergences of an energy
flux become boundary fluxes when integrated over the domain, and so
are omitted for clarity. The rate of change of the components of the
energy density (left side of equations~\ref{eq:energy_keperp}
- \ref{eq:energy_epar}) are given by transfer channels from other
components (right side):
\begin{eqnarray}
\frac{\partial}{\partial t}\left[\frac{1}{2}n_0m_i\left|\frac{\nabla_\perp\phi}{B} + \frac{\nabla_\perp p_i}{n_0 B}\right|^2\right] &=& -\phi\nabla_{||}j_{||} - \phi\nabla\cdot\left[\left(p_e + p_i\right)\nabla\times\frac{\mathbf{b}}{B}\right] \nonumber \\
&& - \frac{p_i}{en_0}\nabla\cdot\left(\mathbf{J}_{||} + \mathbf{J}_d\right) \label{eq:energy_keperp} \\
\frac{\partial}{\partial t}\left(\frac{3}{2}p_i\right) &=& -p_i\nabla\cdot\left(\mathbf{v}_E + \mathbf{b}v_{||i}\right) - W_{ei} \nonumber \\
&& + \frac{p_i}{en_0}\nabla\cdot\left(\mathbf{J}_{||} + \mathbf{J}_d\right) \label{eq:energy_pi} \\
  \frac{\partial}{\partial t}\left(\frac{3}{2}p_e\right) &=& -p_e\nabla\cdot\left(\mathbf{v}_E + \mathbf{b}v_{||e}\right) + W_{ei} \nonumber \\
  && + \left(v_{||i} - v_{||e}\right)F_{ei} \label{eq:energy_pe} \\
  \frac{\partial}{\partial t}\left(\frac{1}{2} m_inv_{||i}^2\right) &=& -v_{||i}\mathbf{b}\cdot\nabla p_i + Z_ienv_{||i}E_{||} - v_{||i}F_{ei} \label{eq:energy_ipar} \\
  \frac{\partial}{\partial t}\left(\frac{1}{2} m_env_{||e}^2\right) &=& -v_{||e}\mathbf{b}\cdot\nabla p_e - env_{||e}E_{||} + v_{||e}F_{ei} \label{eq:energy_epar}
\end{eqnarray}
Each energy transfer channel appears either in two equations with opposite signs, or
combines with other terms to form a full divergence. For example the Alfv\`en wave involves
terms in equations~\ref{eq:energy_keperp}, \ref{eq:energy_ipar} and \ref{eq:energy_epar}.
The electrostatic approximation is used in this model, so $E_{||} = -\partial_{||}\phi$.
\begin{eqnarray}
-\phi\nabla_{||}j_{||} + Z_ienv_{||i}E_{||} - env_{||e}E_{||} &=& -\phi\nabla_{||}j_{||} - j_{||}\partial_{||}\phi \\
&=& \nabla_{||}\left(\phi j_{||}\right)
\end{eqnarray}
so that when integrated over the domain these terms sum to leave only
boundary fluxes.  A similar identity holds for the diamagnetic drift
terms responsible for energy exchange between the $E\times B$ flow
energy, equation \ref{eq:energy_keperp}, and internal energy
equations \ref{eq:energy_pi} and \ref{eq:energy_pe}:
\begin{eqnarray}
-\phi\nabla\cdot\left[p_e\nabla\times\frac{\mathbf{b}}{B}\right] - p_e\nabla\cdot\mathbf{v}_E &=& -\phi\nabla\cdot\left[p_e\nabla\times\frac{\mathbf{b}}{B}\right] - p_e\nabla\times\frac{\mathbf{b}}{B}\cdot\nabla\phi \\
&=& -\nabla\cdot\left[\phi p_e \nabla\times\frac{\mathbf{b}}{B}\right]
\end{eqnarray}
This again has the form of a divergence, and so integrates over the
domain to leave only boundary fluxes. The corresponding terms in the
vorticity and pressure equations are implemented using central
differences, so that the discrete operators for divergence and
gradients obey the above analytic identity.

The term $\frac{p_i}{en_0}\nabla\cdot\left(\mathbf{J}_{||}
+ \mathbf{J}_d\right)$ appearing in perpendicular kinetic energy
(equation~\ref{eq:energy_keperp}) and ion thermal energy
(equation~\ref{eq:energy_pi}) is discussed in section~\ref{apx:ion_energy} above.

\end{document}